\documentclass[usenatbib,useAMS,fleqn,usenames,dvipsnames]{aastex63} 
\usepackage{amsmath}
\usepackage{cancel}
\usepackage{tcolorbox}
\tcbuselibrary{listings,theorems,skins,breakable}
\usepackage{enumerate}

\definecolor{internationalkleinblue}{rgb}{0.0, 0.4, 0.8}
\definecolor{britishracinggreen}{rgb}{0.0, 0.5, 0.3}
\hypersetup{colorlinks=true, 
			citecolor=internationalkleinblue, 
			linkcolor=internationalkleinblue, 
			urlcolor=internationalkleinblue,
			pdftitle={Multiphase Winds},
			pdfauthor={Drummond Fielding}}

\definecolor{dencolor}{HTML}{1b9e77}
\definecolor{momcolor}{HTML}{d95f02}
\definecolor{enecolor}{HTML}{7570b3}
\definecolor{adicolor}{HTML}{E7298A}

\newcommand{\mombox}[1]{\tcboxmath[left=0mm,right=0mm,top=0mm,bottom=0mm,boxrule=0pt,arc=3pt,colback=momcolor!50!white,colframe=white]{#1}}
\newcommand{\denbox}[1]{\tcboxmath[left=0mm,right=0mm,top=0mm,bottom=0mm,boxrule=0pt,arc=3pt,colback=dencolor!50!white,colframe=white]{#1}}
\newcommand{\enebox}[1]{\tcboxmath[left=0mm,right=0mm,top=0mm,bottom=0mm,boxrule=0pt,arc=3pt,colback=enecolor!50!white,colframe=white]{#1}}
\newcommand{\adibox}[1]{\tcboxmath[left=0mm,right=0mm,top=0mm,bottom=0mm,boxrule=0pt,arc=3pt,colback=adicolor!50!white,colframe=white]{#1}}

\newtcbox{\dmebox}[1][]{nobeforeafter,math upper,tcbox raise base,
enhanced,left=0mm,right=0mm,top=0mm,bottom=0mm,boxrule=0pt,arc=3pt,frame hidden,boxrule=0pt,interior style={top color=dencolor!50!white,bottom color=enecolor!50!white,middle color=momcolor!50!white},#1}

\graphicspath{{./}{figures/}}
\received{}
\revised{}
\accepted{}
\submitjournal{ApJ}
\shorttitle{Multiphase Galactic Winds}
\shortauthors{Fielding \& Bryan}

\newcommand{\pushright}[1]{\ifmeasuring@#1\else\omit\hfill$\displaystyle#1$\fi\ignorespaces}

\usepackage{lineno}
\makeatletter
\let\LN@align\align
\let\LN@endalign\endalign
\renewcommand{\align}{\linenomath\LN@align}
\renewcommand{\endalign}{\LN@endalign\endlinenomath}
\let\LN@gather\gather
\let\LN@endgather\endgather
\renewcommand{\gather}{\linenomath\LN@gather}
\renewcommand{\endgather}{\LN@endgather\endlinenomath}
\makeatother

\newcommand{\Msun}{M_{\odot}}
\newcommand{\Msunyr}{M_{\odot} \; {\rm yr}^{-1}}
\newcommand{\kms}{{\rm km \, s}^{-1}}

\newcommand{\mump}{\mu m_{\rm p}}
\newcommand{\kb}{k_{\rm B}}

\newcommand{\Teq}{T_{\rm eq}}

\newcommand{\Ncldot}{\dot{N}_{\rm cl}}

\newcommand{\Tcl}{T_{\rm cl}}
\newcommand{\cscl}{c_{\rm s, cl}}
\newcommand{\ncl}{n_{\rm cl}}
\newcommand{\rhocl}{\rho_{\rm cl}}
\newcommand{\vcl}{v_{\rm cl}}
\newcommand{\vecvcl}{\vec{v_{\rm cl}}}
\newcommand{\vecv}{\vec{v}}
\newcommand{\Mcl}{M_{\rm cl}}
\newcommand{\MZcl}{M_{\rm Z, cl}}
\newcommand{\rcl}{r_{\rm cl}}
\newcommand{\Zcl}{Z_{\rm cl}}
\newcommand{\Zdotcl}{\dot{Z}_{\rm cl}}
\newcommand{\Mdotcl}{\dot{M}_{\rm cl}}
\newcommand{\Mdo}{\dot{M}_{\rm cl,0}}
\newcommand{\MZdotcl}{\dot{M}_{\rm Z, cl}}
\newcommand{\Mdotclgrow}{\dot{M}_{\rm cl,grow}}
\newcommand{\Mdotclloss}{\dot{M}_{\rm cl,loss}}
\newcommand{\pdotcl}{\dot{\varphi}_{\rm cl}}
\newcommand{\vecpdotcltrans}{\dot{\vec{\varphi}}_{\rm cl, trans}}

\newcommand{\vecpdotcldrag}{\dot{\vec{\varphi}}_{\rm cl, drag}}
\newcommand{\Cdrag}{C_{\rm drag}}
\newcommand{\pdotcldrag}{\dot{\varphi}_{\rm cl, drag}}
\newcommand{\pdotdrag}{\dot{p}_{\rm drag}}
\newcommand{\vecpdotclgrav}{\dot{\vec{\varphi}}_{\rm cl, grav}}

\newcommand{\vdotcltrans}{\dot{v}_{\rm cl, trans}}

\newcommand{\vdotcl}{\dot{v}_{\rm cl}}
\newcommand{\edotcl}{\dot{\mathcal{E}}_{\rm cl}}
\newcommand{\edotcool}{\dot{\mathcal{E}}_{\rm cool}}
\newcommand{\edothot}{\dot{\mathcal{E}}_{\rm hot}}

\newcommand{\fcool}{f_{\rm cool}}

\newcommand{\fmix}{f_{\rm mix}}
\newcommand{\fturb}{f_{\rm turb}}

\newcommand{\tmix}{t_{\rm mix}}

\newcommand{\vin}{v_{\rm in}}

\newcommand{\vturb}{v_{\rm turb}}

\newcommand{\vturbcl}{v_{\rm turb,cl}}

\newcommand{\tcoolmix}{t_{\rm cool,mix}}
\newcommand{\tcool}{t_{\rm cool}}
\newcommand{\tcoolwind}{t_{\rm cool, wind}}
\newcommand{\tcl}{\tau_{\rm cool}}
\newcommand{\vrel}{v_{\rm rel}}
\newcommand{\Mach}{\mathcal{M}}

\newcommand{\rhodot}{\dot{\rho}}
\newcommand{\rhodotp}{\dot{\rho}_+}
\newcommand{\rhodotm}{\dot{\rho}_-}
\newcommand{\rhoZ}{\rho_Z}
\newcommand{\rhoZdot}{\dot{\rho}_Z}
\newcommand{\Zdot}{\dot{Z}}
\newcommand{\pdot}{\dot{p}}
\newcommand{\edot}{\dot{\varepsilon}}
\newcommand{\vB}{v_\mathcal{B}}
\newcommand{\vBcl}{v_{\mathcal{B},\rm cl}}
\newcommand{\Mdotcold}{\dot{M}_{\rm cold}}
\newcommand{\Mdothot}{\dot{M}_{\rm hot}}
\newcommand{\Mdotwind}{\dot{M}_{\rm wind}}

\newcommand{\Across}{A_{\rm cross}}
\newcommand{\Acool}{A_{\rm cool}}

\newcommand{\Omwind}{\Omega_{\rm wind}}
\newcommand{\rstar}{r_\star}

\newcommand{\Mdotstar}{\dot{M}_\star}
\newcommand{\Edotstar}{\dot{E}_\star}
\newcommand{\mstar}{m_\star}
\newcommand{\Mejecta}{M_{\rm ejecta}}
\newcommand{\etaMtot}{\eta_{\rm M, tot}}
\newcommand{\etaMcold}{\eta_{\rm M, cold}}
\newcommand{\etaM}{\eta_{\rm M, hot}}
\newcommand{\etaE}{\eta_{\rm E}}

\newcommand{\ESN}{{E}_{\rm SN}}
\newcommand{\Edotwind}{\dot{E}_{\rm wind}}

\newcommand{\tcc}{t_{\rm cc}}
\newcommand{\tdrag}{t_{\rm drag}}

\newcommand{\cshot}{c_{\rm s,hot}}
\newcommand{\vesc}{v_{\rm esc}}
\newcommand{\vc}{v_{\rm c}}
\newcommand{\cs}{c_{\rm s}}

\newcommand{\dr}[1]{\frac{\partial #1}{\partial r}}

\newcommand{\dloglogr}[1]{\frac{\partial\log #1}{\partial \log r}}

\newcommand{\lr}[1]{\left( #1 \right)}

\newcommand{\SigmaSFR}{\Sigma_{\rm SFR}}
\newcommand{\Sigmagas}{\Sigma_{\rm gas}}

\newcommand{\eg}{e.g.,}

\defcitealias{Fielding:2020}{F20}
\defcitealias{Gronke:2018}{GO18}
\defcitealias{Gronke:2020}{GO20}
\defcitealias{Tan:2021}{TOG21}
\defcitealias{ChevalierClegg:1985}{CC85}

\begin{document}

\title{The Structure of Multiphase Galactic Winds}

\correspondingauthor{Drummond B. Fielding}
\email{drummondfielding@gmail.com}

\author[0000-0003-3806-8548]{Drummond B. Fielding}
\affiliation{Center for Computational Astrophysics, Flatiron Institute, 162 5th Ave, New York, NY 10010, USA}

\author[0000-0003-2630-9228]{Greg L. Bryan}
\affiliation{Department of Astronomy, Columbia University, 550 W 120th Street, New York, NY 10027, USA}
\affiliation{Center for Computational Astrophysics, Flatiron Institute, 162 5th Ave, New York, NY 10010, USA}

\begin{abstract}
We present a novel analytic framework to model the steady-state structure of multiphase galactic winds comprised of a hot, volume-filling component and a cold, clumpy component. We first derive general expressions for the structure of the hot phase for arbitrary mass, momentum, and energy sources terms. Next, informed by recent simulations, we parameterize the cloud-wind mass transfer rates, which are set by the competition between turbulent mixing and radiative cooling. This enables us to cast the cloud-wind interaction as a source term for the hot phase and thereby simultaneously solve for the evolution of both phases fully accounting for their bidirectional influence. With this model, we explore the nature of galactic winds over a broad range of conditions. We find that: (i) with realistic parameter choices, we naturally produce a hot, low-density wind that transports energy while entraining a significant flux of cold clouds, (ii) mixing dominates the cold cloud acceleration and decelerates the hot wind, (iii) during mixing thermalization of relative kinetic energy provides significant heating, (iv) systems with low hot-phase mass loading factors and/or star formation rates can sustain higher initial cold phase mass loading factors, but the clouds are quickly shredded, and (v) systems with large hot-phase mass loading factors and/or star formation rates cannot sustain large initial cold-phase mass loading factors, but the clouds tend to grow with radius. Our results highlight the necessity of accounting for the multiphase structure of galactic winds, both physically and observationally, and have important implications for feedback in galactic systems.
\end{abstract}


\section{Introduction} \label{sec:intro}
In essence, galaxy formation can be seen as a competition of inflows and outflows to regulate the fuel supply for star formation and black hole growth. The importance of galactic winds has long been appreciated owing to the fact that galactic accretion brought on by radiative cooling of virial shock heated gas in the circumgalactic medium (CGM), if unchecked, would lead to an overproduction of stars relative to what is observed \citep[\eg][]{White:Rees:1978, Dekel:1986, White:1991}. Galactic winds limit the formation of stars both by heating CGM material and thus \emph{preventing} inflows from reaching the galaxy, and by \emph{ejecting} gas out of the galaxy into the CGM and beyond. The source powering these outflows is tied to the \emph{feedback} from star formation (SF) and/or active galactic nuclei (AGN). Underlying these feedback processes is a complex interplay of physical mechanisms that interact over a huge range of scales making the development of a coherent theory for galactic winds as daunting as it is important.

Traditional theoretical work has treated galactic winds as being comprised solely of hot, low density gas \citep[\eg ][hereafter \citetalias{ChevalierClegg:1985}]{ChevalierClegg:1985}. Observations of galactic winds, however, generally probe much colder and denser gas than is accounted for in these models \citep[\eg][]{Heckman:1990}. A comprehensive theory that takes into account the highly multiphase nature of galactic winds is necessary to understand their role in galaxy evolution, and to extract as much information as possible from complex observations. That motivates us to work towards a new model for the co-evolution of a hot wind and the cold clouds embedded within. 

\subsection{Observing Galactic Winds}
Ample observational evidence helps inform our understanding of the nature and inner workings of galactic winds. Galactic winds are observed in the local universe \citep[\eg][]{Lehnert:1996, Martin:1999}, as well as in systems at high redshifts of $z \gtrsim 2$ \citep[\eg][]{Pettini:2001, Shapley:2003}. These observations have made it clear that galactic winds are a ubiquitous feature of star-forming galaxies \citep[][]{Weiner:2009,Rubin:2014}. Any successful model of galactic winds must confront the known observational trends. For example, wind velocities tend to be higher in systems with higher star formation rates (SFRs) \citep[\eg][]{Martin:2005,Heckman:2015}, and winds from lower stellar mass galaxies with lower SFRs have larger mass outflow rates relative to their SFR (as characterised by the mass loading factor $\eta_{\rm M} \equiv \dot{M}_{\rm wind}/\dot{M}_\star$) than their larger counterparts \cite[\eg][]{Newman:2012,Chisholm:2017}. 

One of the most important observational findings about galactic winds is their multiphase nature. Galactic winds are observed in both emission and absorption using probes that trace gas at temperatures ranging from $\sim 10$ K all the way up to $\gtrsim 10^7$ K. The coldest phase of outflows $\sim 10-100$ K is traced by neutral hydrogen H\textsc{i} \citep[\eg][]{Walter:2002, Martini:2018}, molecules, such as, CO \citep[\eg][]{Bolatto:2013,Cicone:2014,Leroy:2015}, and neutral metals, such as NaD, O\textsc{i}, and Mg\textsc{i} \citep{Heckman:2000,Rupke:2005a,Chen:2010}. Somewhat warmer $\sim 10^{4-5}$ K outflowing gas is observed in photoionized metals, including Mg\textsc{ii}, O\textsc{iii}, and Si\textsc{iv} \citep[\eg][]{Martin:2009,Rubin:2011,Nielsen:2015,Chisholm:2017}, and H$\alpha$ \citep[\eg][]{McKeith:1995,Westmoquette:2009}. Yet hotter phases of outflows $\sim 10^{5-6}$ K are traced by more ionized metal species such as O\textsc{vi} \citep[\eg][]{Steidel:2010,Kacprzak:2015,Chisholm:2018,Ashley:2020}. X-ray emission is used to measure the hottest phase $\sim 10^7$ K of galactic winds \citep[\eg][]{Ptak:1997, Strickland:2009, Lopez:2020, Hodges-Kluck:2020}. Although rare to simultaneously have measurements from all phases in all but the most optimal nearby sources (\eg\ M82 and NGC253), the evidence is clear that there is both hot and cold material emanating from galaxies at velocities of 100s to 1000s of km/s. Despite the wealth of observational data, a precise determination of the properties of winds and what drives them remains elusive because of inherent difficulties in disentangling the location and physical nature of the observed gas. Upcoming emission observations with KCWI and MUSE provide a promising path forward to greatly improve our understanding galactic wind as has been recently demonstrated observationally \citep[][]{Hayes:2016,Rupke:2019,Burchett:2021} and theoretically \cite[\eg][]{Nelson:2021}.

\subsection{Modeling Galactic Winds}
From a theoretical standpoint, the two major goals related to galactic winds are to understand (i) the role of winds in galaxy formation and evolution, and (ii) the properties and driving mechanisms of the winds. However, given the huge spatial and temporal scales inherent in modeling galaxies on cosmological scales it is infeasible to simultaneously capture the detailed and fundamentally small scale processes occurring in a galaxy's interstellar medium (ISM) and in the wind \citep[see][and references therein]{Naab:Ostriker:2017}. As a result parameterized \textit{subgrid} models are commonly adopted to determine when stars and black holes form, and to control how their feedback impacts the galaxy and launches winds. These subgrid models come in a variety of forms and approaches and are generally calibrated to reproduce specific observations \citep[\eg][]{Vogelsberger:2013, Crain:2015, Pillepich:2018}, and/or higher resolution simulations \citep[\eg][]{Hopkins:2014,Dave:2019}. Using subgrid models cosmological simulations and semi-analytic models demonstrated that the inclusion of SF and AGN feedback and the resulting galactic winds are essential to match observed statistical properties of galaxies, such as the stellar mass function and mass-metallicity relation \citep[see][and references therein]{Somerville:Dave:2015}. Beyond statistical properties, cosmological simulations also provide valuable insights into the processes shaping galaxy evolution and the structure of material in and around galaxies \citep[\eg][]{Keres:2005,Oppenheimer:2008,Nelson:2013,Stern:2020,Stern:2021,Mitchell:2020,Fielding:2020b}. Detailed comparisons with observations are, however, fraught due to the reliance on subgrid models and insufficient spatial resolutions to reliably capture the colder phases. In fact, the vast majority of subgrid feedback models do not account for the multiphase nature of outflows in any way \citep[with the notable exception of the PhEW model][which is closely related to what we present in this paper]{Huang:2020,Huang:2021}. Despite this, in recent high resolution cosmological simulations, such as TNG50 and FIRE, multiphase structure develops self-consistently in the winds as they propagate into the surrounding medium \citep{Nelson:2019, Mitchell:2020, Pandya:2021}.

High resolution individual galaxy simulations and ISM patch simulations can more clearly illuminate the detailed properties of galactic winds and launching mechanisms.
Winds driven by SF---the primary focus of this work---are powered by the feedback associated with young stars, which includes stellar winds and radiation, but is dominated by energy released by core-collapse supernovae (SNe). To study the launching mechanism it is, therefore, necessary to resolve this energy injection and its interaction with the surrounding medium. Simulations of this sort require pc-scale resolution and thus can only be run on domains of limited size. Simulations of $\sim$kpc patches of the ISM run have been run to this end with a varying degree of physical realism \citep[\eg][]{deAvillez:2000,Joung:2006,Hill:2012}. More controlled and idealized experiments have demonstrated, among other things, that the strength of outflows is sensitive to SN placement \citep[\eg][]{Creasey:2013, Martizzi:2016, Girichidis:2016a, Li:2017}, and the degree of spatio-temporal SN clustering \citep[\eg][]{Kim:2017, Fielding:2018}.  Among the most realistic of the ISM patch simulations are the TIGRESS simulations, which self-consistently form stars and drive multiphase winds (\citealt{Kim:2018}, see also \citealt{Gatto:2017}). A recent analysis of these simulations demonstrated that across the wide range of conditions they simulated the hot phase ($\gtrsim 10^6$ K) carries the majority of the energy, while the cold phase ($\sim 10^4$ K) carries most of the mass, with the cold phase mass loading factor $\etaMcold \equiv \Mdotcold / \Mdotstar$ increasing as the star formation rate surface density decreases \citep{Kim:2020a,Kim:2020b}. Broadly, these trends are corroborated by quite disparate recent ISM patch simulations \citep[see][for a compilation of recent work]{Li:2020a}, and are strikingly different from what are generally used in cosmological simulation subgrid models.

Individual galaxy simulations provide an ideal complement to the ISM patch simulations because they make up for the inherent deficiencies of the patch simulations related to the lack of room for the winds to expand into \citep{Fielding:2017b}. This generally comes at the expense of the number of physical processes included and/or only being able to simulate dwarf galaxies \citep[\eg][]{Suchkov:1994, Tanner:2016, Smith:2018, Vijayan:2018, Emerick:2018, Schneider:2018, Hu:2019}. \cite{Schneider:2020} simulate an M82-like galaxy with 5 pc resolution in a 10 kpc volume, which is ideally suited to capture both the launching of the wind and the detailed phase structure during the subsequent expansion. At the base of the wind, near the galaxy, they find multiphase mass and energy fluxes in close agreement with ISM patch simulations. However, they demonstrate that farther out in the wind the cold phase is predominantly in the form of embedded clouds that are gradually shredded and thereby enhance the mass loading of the hot phase. Beyond the obvious implications for galactic wind observations this striking behavior of the change in the partitioning of the mass, momentum, and energy flux between the phases will have a profound impact on the overall galactic feedback cycle \citep[\eg][]{Angles-Alcazar:2017,Fielding:2017a}, and the (observable) structure of the CGM \citep[\eg][]{Hummels:2013,Fielding:2020b}.

Analytic models for galactic winds provide a useful bridge between simulation and observation. They can provide an intuitive understanding of the qualitative behavior of what is observed in the real universe and are useful to disentangle complex and expensive simulations. 
There is a long history of work developing models for the steady state structure of galactic winds \citep[\eg][]{Mathews:1971, Morita:1982}. The \citetalias{ChevalierClegg:1985} model represents the simplest case in which the wind is powered by the injection of mass and energy in a finite region. The resulting wind is subsonic within the injection region and smoothly transitions to supersonic beyond. There have been numerous extensions to this model to include the effects of radiative cooling and gravity \citep[\eg][]{Wang:1995a, Silich:2003, Silich:2011, Thompson:2016}, more extended sources of mass and energy injection \citep[\eg][]{Bustard:2016}, and sources of momentum injection \citep[\eg][]{SharmaNath:2013}. These models have demonstrated the richness of behavior beyond the simple \citetalias{ChevalierClegg:1985} solutions, however, they are all fundamentally single-phase, in that at a given radius the material is all at a single temperature. Other models have attempted to account for the multiphase nature of galactic winds by including a cold embedded cloud phase that exchanges mass, momentum, and energy with the hot, volume filling phase \citep[\eg][]{Cowie:1981,Suchkov:1996,Zhang:2017}. The behavior of and conclusions drawn from these multiphase galactic wind models are, however, sensitive to the nature of the cloud-wind interaction and the corresponding exchange rates, which are far less well understood than, for example, gravity or radiative cooling. Notably, to our knowledge in all existing analytic multiphase galactic wind models, the cold clouds can only lose mass. 

\subsection{Cold Clouds in a Hot Wind}

The interaction of a cold cloud moving relative to a hot wind has been studied extensively in a wide range of contexts using what are known as ``cloud-crushing'' simulations \citep[\eg][]{Klein:1994, Xu:1995}. These simulations are varied in their details, but generally involve hitting a cold cloud with a hot wind. A seminal result of these simulations is that the clouds tend to be destroyed by Kelvin-Helmholtz (KH) and Rayleigh-Taylor (RT) instabilities on the cloud crushing timescale $\tcc \equiv \chi^{1/2} \rcl / \vrel$, where $\rcl$ is the cloud radius, $\vrel$ is the relative velocity of the cloud and the wind, and $\chi \equiv \rhocl/\rho$ is the density contrast between the cloud and wind. At the same time clouds are accelerated by drag/ram pressure, which acts on the drag timescale $\tdrag \equiv \chi \rcl / \vrel$. In most galactic wind environments $\chi \gg 1$, so $\tdrag \gg \tcc$ and the clouds are destroyed prior to being entrained by the wind. This is problematic when considering the abundance of observational evidence for fast moving cold clouds in galactic winds \citep{Zhang:2017}. Many groups have investigated how the cloud-wind interaction is modified by the presence of additional physical processes, including radiative cooling \citep[\eg][]{Mellema:2003, Cooper:2009, Scannapieco:2015}, conduction \citep[\eg][]{Marcolini:2005,Orlando:2005,Bruggen:2016}, magnetic fields \citep[\eg][]{MacLow:1994, McCourt:2015, Schneider:2017, Cottle:2020}, and/or cosmic rays \citep[\eg][]{Wiener:2019,Bruggen:2020}. While conduction, magnetic fields, and cosmic rays can all quantitatively change the destruction and acceleration rates only radiative cooling can qualitatively change the nature of the cloud-wind interaction by causing the cloud to gain mass under certain conditions, rather than lose mass \citep[\eg][]{Marinacci:2010,Armillotta:2016}. \cite{Gronke:2018} (hereafter \citetalias{Gronke:2018}) provided an influential and intuitive picture for the impact of cooling based on the rough criterion that clouds will grow when the cooling time of material in the mixing layer between the cold cloud and the hot wind is shorter than the cloud crushing time, i.e., $\tcoolmix < \tcc$. This translates into a cloud size criterion in which large clouds grow because they cool more rapidly than they are shredded. This sparked great debate into the exact nature of this criterion and how it is modified by additional physical processes, which has yet to be conclusively answered \citep[\eg][]{Gronke:2020,Li:2020,Sparre:2020,BandaBarragan:2020,Kanjilal:2021,Abruzzo:2021}.

Additionally, even higher resolution focused simulations that isolate the turbulent radiative mixing layers (TRMLs) which separate the phases and mediate their interactions have begun to further refine this picture \citep[\eg][]{Ji:2019, Mandelker:2020}. In particular, these simulations have opened a window into the fundamental nature of TRMLs and have lead to the development of comprehensive models for the mass, momentum, and energy transfer rates based on the fractal nature of the interface \citep[][hereafter \citetalias{Fielding:2020}]{Fielding:2020} and the parallels to turbulent combustion \citep[][hereafter \citetalias{Tan:2021}]{Tan:2021}. On an intuitive level the picture that has emerged is one in which the relative motion of the cloud and wind drives turbulence, which in turn leads to mixing of the phases and populates the intermediate temperature regime. The intermediate temperature $\sim 10^5$ K material radiatively cools on very short timescales \citep[\eg][]{SutherlandDoptia:1993}, which leads to the advection of hot, high enthalpy material into the mixing layer to replace the energy lost to radiation. If the mixing time $\tmix$, or eddy turn over time, of the turbulence in the mixing layer is longer than the cooling time then the interface between the phases will be a fractal, highly corrugated surface. On the other hand, if the mixing time is shorter than the cooling time there is a thicker, more gradual transition between the phases. The rate at which hot material is advected into the mixing layer and condenses on to the cold phase, bringing mass, momentum, and energy, changes in these two limits, with the cold phase growth rate scaling with $\xi^{1/4}$ in the rapid limit (for which $\xi>1$), or as $\xi^{1/2}$ in the slow limit ($\xi<1$), where $\xi = \tmix / \tcoolmix$ \citepalias{Fielding:2020, Tan:2021}. The thickness and corresponding phase distribution of the layer also changes in these two regimes \citep{Tan:2021b}.

\subsection{Our approach}

\begin{figure*}
\centering
\includegraphics[width=\textwidth]{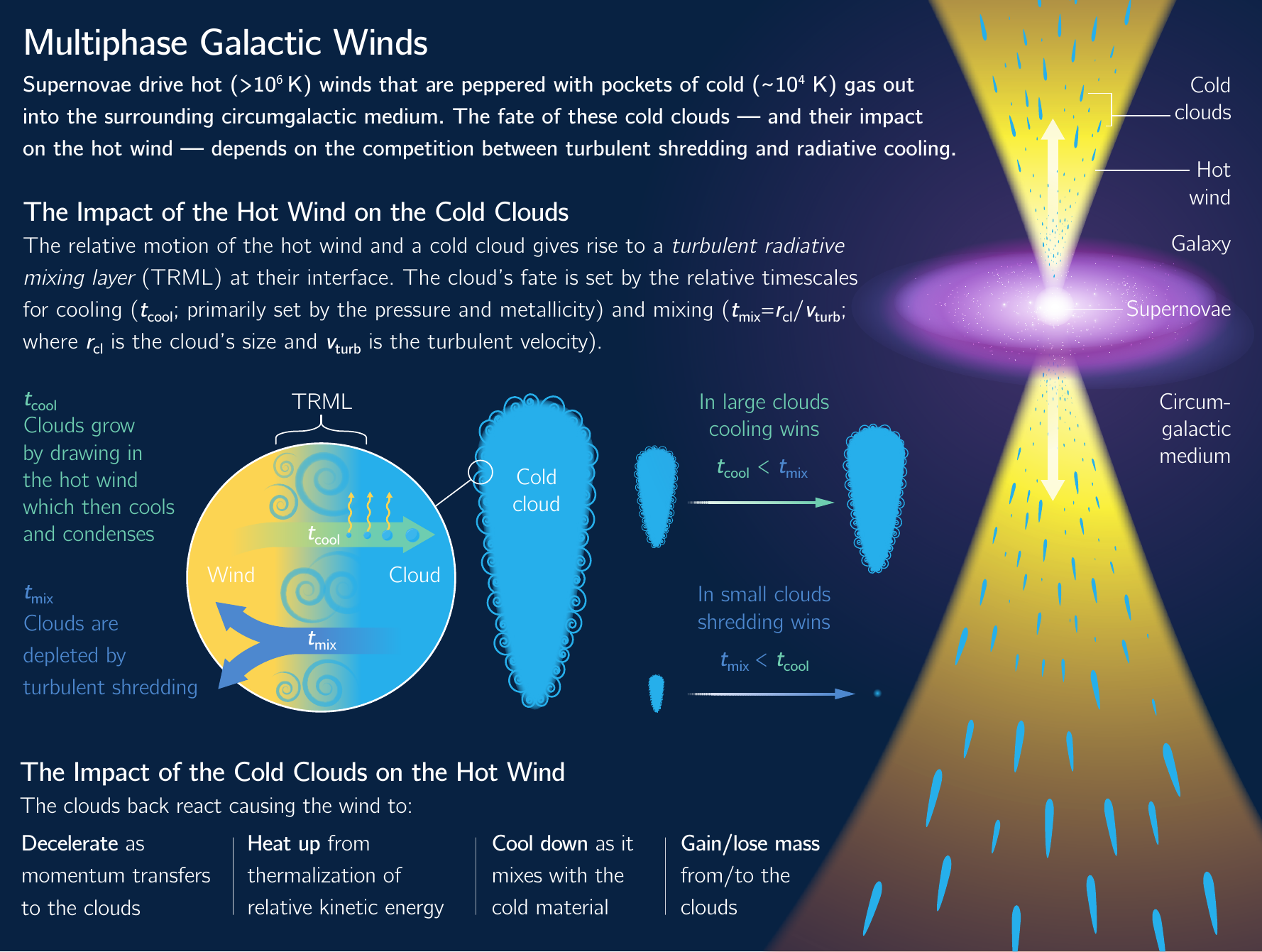} 
\caption{Schematic summary of our multiphase galactic wind model in which cold clouds and a hot volume-filling phase co-evolve with interactions mediated by the turbulent radiative mixing layers that separate them.}
\label{fig:Schematic}
\end{figure*}

To date, the vast majority of studies of multiphase galactic winds have focused on how the hot wind impacts the cold clouds. With a deeper understanding of this interaction based on the new turbulent radiative mixing layer models we are poised to address the relatively unexplored, yet crucial, back-reaction of the cold clouds on the hot wind. An underlying assumption of most cloud wind studies is that the wind represents an effectively infinite reservoir of mass, energy, and momentum. Multiphase galactic winds, however, often fall in the regime where the mass, energy, and momentum that is transferred between the phases can significantly impact the bulk structure and dynamics of the galactic wind. Therefore, in order to (i) interpret galactic wind observations, (ii) understand the results of galaxy formation simulations, and (iii) develop a comprehensive model of the role of galactic winds (and more generally multiphase flows) in galaxy evolution we must understand the co-evolution of the hot and cold phase of galactic winds. To this end we now lay out the basics of just such a model along with promising preliminary results. The current form of our model allows us to calculate spherically symmetric, steady-state profiles for multiphase galactic winds, however, in an upcoming work we will use the same underlying model as a subgrid model for isolated and cosmological galaxy formation simulations.

\autoref{fig:Schematic} lays out the basic premise and summarizes the main components of our new model. The galactic wind is comprised of a hot, volume-filling component and a cold component in the form of embedded clouds. As these two components co-evolve they exchange mass, momentum, and energy which modifies the structure of the wind relative to single-component wind models and causes the cold clouds to gain or lose mass while being accelerated. The two phases in our model exchange mass, momentum, and energy as mediated by the turbulent radiative mixing layers (TRML) that are seen in high resolution simulations. The cooling in these mixing layers can, in many cases, dominate the total cooling of the multiphase wind. 

In the following sections we incrementally build up the general framework of our model. First, in \autoref{sec:volumefilling}, we introduce the equations governing the hot, volume-filling phase of the galactic wind. This starts in \autoref{sec:review_limit_cases} and \autoref{sec:hotphase_mass_loss} with a somewhat pedagogical review of past work and simple limiting cases. In \autoref{sec:source_terms_relative_velocity} we present the equations describing the structure of the hot phase of a wind under the influence of source terms with a relative velocity, which forms the basis of our model. We then describe a simple model for how an individual embedded cloud evolves due to its interaction with the hot, volume-filling phase in \autoref{sec:cloudevo}. \autoref{sec:single_cloud} contains the general form for the mass, momentum, and energy transfer rates for an arbitrary mass transfer model, and in \autoref{sec:mass_transfer_model} we present the particular mass transfer model we adopt throughout the rest of the paper. In \autoref{sec:coevolution} we introduce the full multiphase galactic wind model in which the evolution of the hot phase and the clouds are coupled together. In \autoref{sec:low_eta} and \autoref{sec:high_eta} we explore the nature of the multiphase wind solutions given by our model for a range of systems and dissect several limiting solutions in detail to determine the exact processes responsible for the wind behavior. The general trends of the solutions with key parameters such as the number of cold clouds, and the star formation rate (SFR) are presented in \autoref{sec:general_trends}. In \autoref{sec:discussion} we discuss the implications and connection of our model to simulations and observations of galactic winds, as well as the deficiencies of our model and how we plan to extend it in the future. We summarize our model and the key findings in \autoref{sec:summary}

\section{Structure of the Hot, Volume-Filling Phase of Galactic Winds} \label{sec:volumefilling}

In this section we consider the steady-state structure of the hot, volume-filling phase of a galactic wind. In the case of a single-phase wind this describes the entire structure of the wind, but in later sections we will combine this with an additional embedded cloud phase of the wind. We work under the assumption that the winds subtend a solid angle $\Omwind$ and within that solid angle they are spherically symmetric. In the case of dwarf galaxies or extremely powerful starburst the wind may occupy the full $4 \pi$, while in more massive galaxies the wind may cover a significantly smaller fraction of the sky and take the form of a bi-conical outflow along the minor axis of the galaxy. The steady-state equations for mass, momentum, and energy conservation are
\begin{align}
    &\frac{1}{r^2} \dr{} \lr{r^2 \rho v} = \rhodot \\
    &\frac{1}{r^2} \dr{} \lr{r^2 \rho v^2} + \dr{P} = -\rho \frac{\vc^2}{r} + \pdot \\
    &\frac{1}{r^2} \dr{} \lr{r^2 \rho v \lr{ \frac{1}{2} v^2 + \frac{\gamma}{\gamma{-}1} \frac{P}{\rho} - \frac{1}{2} \vesc^2 } } =  \edot - \mathcal{L},
\end{align}
where the circular velocity, $\vc^2 = r\, d\Phi/dr$, and escape velocity, $\vesc^2 = -2 \Phi$, come from the gravitational potential of the galaxy and dark matter halo. The effect of radiative cooling and heating are encompassed by $\mathcal{L} = n^2 \Lambda - n \Gamma$. We have included three less common terms in these expressions, $\rhodot$, $\pdot$, and $\edot$ that are source (or sink) terms for mass, momentum, and energy. The energy source term $\edot$  includes the change in energy from the momentum source term.  These allow us to write down the familiar form of the velocity and entropy equations
\begin{align}
    &v \dr{v} = - \frac{1}{\rho} \dr{P} - \frac{\vc^2}{r} + \frac{\pdot - \rhodot v}{\rho} \\
    &\rho T v \dr{s} = \frac{P v}{\gamma - 1}\dr{\log K} = \frac{v}{\gamma - 1}\left( \dr{P} - \frac{\gamma P}{\rho}\dr{\rho} \right) = \edot - \mathcal{L} - \pdot v - \rhodot (\vB^2 -v^2) \label{eq:lagranian} \\
    &\quad  \text{where }  \vB^2 \equiv \frac{1}{2} v^2 + \frac{\gamma}{\gamma{-}1} \frac{P}{\rho} - \frac{1}{2} \vesc^2,\; K \equiv \frac{P}{\rho^\gamma}, \text{ and } s = (\gamma-1)\frac{\kb}{\mump} \log\lr{\frac{P}{\rho^\gamma}}. \nonumber
\end{align}
These equations can be expanded and rearranged to give expressions for the change in $v$, $\rho$, and $P$ with radius. We find the most useful form of these expressions to be 
\begin{align}
    \dloglogr{v} &= \frac{1}{1-\Mach^{-2}} \left( \frac{2}{\Mach^2} - \frac{\vc^2}{v^2} - \frac{\rhodot/\rho}{v/r} \left( \frac{\gamma+1}{2} - \gamma \frac{\pdot}{\rhodot v} + (\gamma-1) \frac{\edot - \mathcal{L}}{\rhodot v^2} + \frac{\gamma-1}{2}\frac{\vesc^2}{v^2} \right)\right) \label{eq:velocity_gradient} \\
    \dloglogr{\rho} &= \frac{1}{1-\Mach^{-2}} \left(-2 + \frac{\vc^2}{v^2} + \frac{\rhodot / \rho}{v/r} \left( \frac{\gamma+3}{2} - \gamma \frac{\pdot}{\rhodot v} + (\gamma-1)\frac{\edot - \mathcal{L}}{\rhodot v^2} - \frac{1}{\Mach^2} + \frac{\gamma-1}{2} \frac{\vesc^2}{ v^2} \right) \right) \label{eq:density_gradient} \\
    \dloglogr{P} &= \frac{\gamma}{1-\Mach^{-2}} \left(-2 + \frac{\vc^2}{v^2} + \frac{\rhodot/\rho}{v/r} \left(1 + \frac{\gamma-1}{2}\Mach^2 \left( 1 +\frac{\vesc^2}{v^2}\right)  - \frac{\pdot}{\rhodot v} + \left((\gamma-1)\Mach^2 \right)  \frac{\edot - \mathcal{L} - \pdot v}{\rhodot v^2} \right)\right), \label{eq:pressure_gradient} 
\end{align}
where $\cs^2 = \gamma P / \rho$, and $\Mach = v/\cs$. These expressions are not readily interpretable given how many terms there are in each equation. We, therefore, start by considering the simplest limiting cases and incrementally build up in complexity.

\subsection{Review of standard limiting cases} \label{sec:review_limit_cases}

The simplest case is one in which (i) $v \gg \vc$ and $\vesc$ (we will refer to this as a super-virial flow), so we can neglect gravity,  (ii) there is negligible radiative cooling and heating $\mathcal{L} = 0$, and (iii) there are no additional source terms. In this limit we recover the standard adiabatic wind equations that have been used extensively to model galactic winds. If the flow is highly supersonic ($\Mach \gg 1$) then the solution is given by
\begin{subequations}
\label{eq:simple_supersonic}
\begin{align}
    \dloglogr{v} &= 0\\
    \dloglogr{\rho} &= -2 \\ 
    \dloglogr{P} &= - 2 \gamma. 
\end{align} 
\end{subequations}
As a result this supersonic adiabatic wind solution has $T \propto r^{-2(\gamma -1)}$ and $K \propto$ constant. In the opposite limit in which the flow is highly subsonic ($\Mach \ll 1$) the solutions satisfy $v\propto r^{-2}$ and $\rho,P \propto$ constant. There is no smooth solution that goes from subsonic to supersonic at finite radius in this simple limiting case. However, in almost all cases we will be interested solely in supersonic winds and so the limiting behavior in \autoref{eq:simple_supersonic} provides a useful reference solution.

With this intuition in hand we can now move onto the case in which there is a source of mass and energy (i.e., $\rhodot, \edot > 0$) in a finite region of space. This forms the basis for the seminal \citetalias{ChevalierClegg:1985} galactic wind model in which mass and energy are added by SNe within the star forming region of a galaxy. The \citetalias{ChevalierClegg:1985} model neglects gravity and cooling, but is a useful zeroth-order model for galactic winds. Given the utility of the \citetalias{ChevalierClegg:1985} model we adopt a similar set of assumptions for relating the mass and energy source terms to the star formation, and will reproduce, then extend, their findings.

For galactic winds driven by SNe (and other forms of feedback associated with star formation) it is useful to relate mass and energy injection rates to the star formation rate. The mass injection rate is commonly parameterized in terms of the mass loading factor $\etaM \equiv \Mdotwind / \Mdotstar$, where $\Mdotwind$ is the total mass flux (in steady state this is necessarily equivalent to the mass injection rate) and $\Mdotstar$ is the star formation rate. We specifically use the `hot' subscript in this definition of the mass loading factor because in subsequent sections it will be necessary to distinguish between the mass flux and mass loading factor in the hot and cold phases. Similarly, we can relate the total energy injection rate $\Edotwind$ to the energy available from star formation $\Edotstar$ by introducing the energy loading factor $\etaE \equiv \Edotwind/\Edotstar$. SNe are the most energetic sources of star formation feedback, so we make the common assumption that they dominate the energy available from star formation. The energy injected by SNe can be related to the star formation rate by assuming that every SN deposits $\ESN = 10^{51}$ erg and that one SN happens for every $\mstar = 100~\Msun$ that is formed. In which case $\Edotwind = \etaE (\ESN / \mstar) \Mdotstar$. It is crucial to appreciate that none of our findings depend on these assumptions, however, they allow us to specify the conditions at the base of the wind using $\Mdotstar$, $\etaM$, and $\etaE$, instead of $v$, $\rho$, and $P$, which makes the the connection to galaxy formation clearer. It is trivial to modify any of these assumptions specifying how the mass and energy injection rates relate to galactic properties (e.g., for AGN driven winds). 

Following \citetalias{ChevalierClegg:1985} we assume that the mass and energy from the star formation feedback are added uniformly within the star forming region which extends out to $\rstar$. The mass source term, therefore, is $\rhodot = \etaM \Mdotstar / (4 \pi \rstar^3 /3)$ and the energy source term is $\edot = \etaE (\ESN / \mstar) \Mdotstar / (4 \pi \rstar^3 /3)$. In this case the wind is subsonic out to $\rstar$, at which point it smoothly transitions to being supersonic. The wind properties at the sonic point, $v_0,~\rho_0,~P_0$, can be found using the fact that $\Mdotwind = \Omwind \rstar^2 \rho_0 v_0 \text{ and } \Edotwind = \Mdotwind (v_0^2/2 + \gamma/(\gamma -1) P/\rho)$, which implies that 
\begin{subequations}
\label{eq:sonic_radius_values}
\begin{align}
    v_0 &= \lr{\frac{\Edotwind}{2 \Mdotwind}}^{1/2} = 500\:{\rm km\,s}^{-1}\:\etaE^{1/2}\etaM^{-1/2}  \\
    \rho_0 &= \frac{2^{1/2} \Mdotwind^{3/2}}{\Omwind \rstar^2 \Edotwind^{1/2}} = 1.18 \times 10^{-25} \: {\rm g\,cm}^{-3}\:\etaE^{-1/2}\etaM^{3/2}\lr{\frac{\Mdotstar}{\Msun {\rm yr}^{-1}}}\lr{\frac{4 \pi}{\Omwind}} \lr{\frac{300 \: {\rm pc}}{\rstar}}^2    \\
    P_0/\kb &= \frac{\Edotwind^{1/2} \Mdotwind^{1/2}}{\gamma 2^{1/2} \Omwind \rstar^2} = 1.28 \times 10^{6} \:{\rm K\,cm}^{-3}\:\etaE^{1/2}\etaM^{1/2}\lr{\frac{\Mdotstar}{\Msun {\rm yr}^{-1}}} \lr{\frac{4 \pi}{\Omwind}}\lr{\frac{300 \: {\rm pc}}{\rstar}}^2
\end{align}
\end{subequations}
where we have adopted $\gamma = 5/3$ as we will for the remainder of this paper. These values of the velocity, density, and pressure at the sonic point serve as the (hot wind) initial conditions for all numerical integrations. To find the subsonic solution we integrate inward from $\rstar$, and to find the supersonic solution we integrate outward from $\rstar$.

\begin{figure*}
\centering
\includegraphics[width=\textwidth]{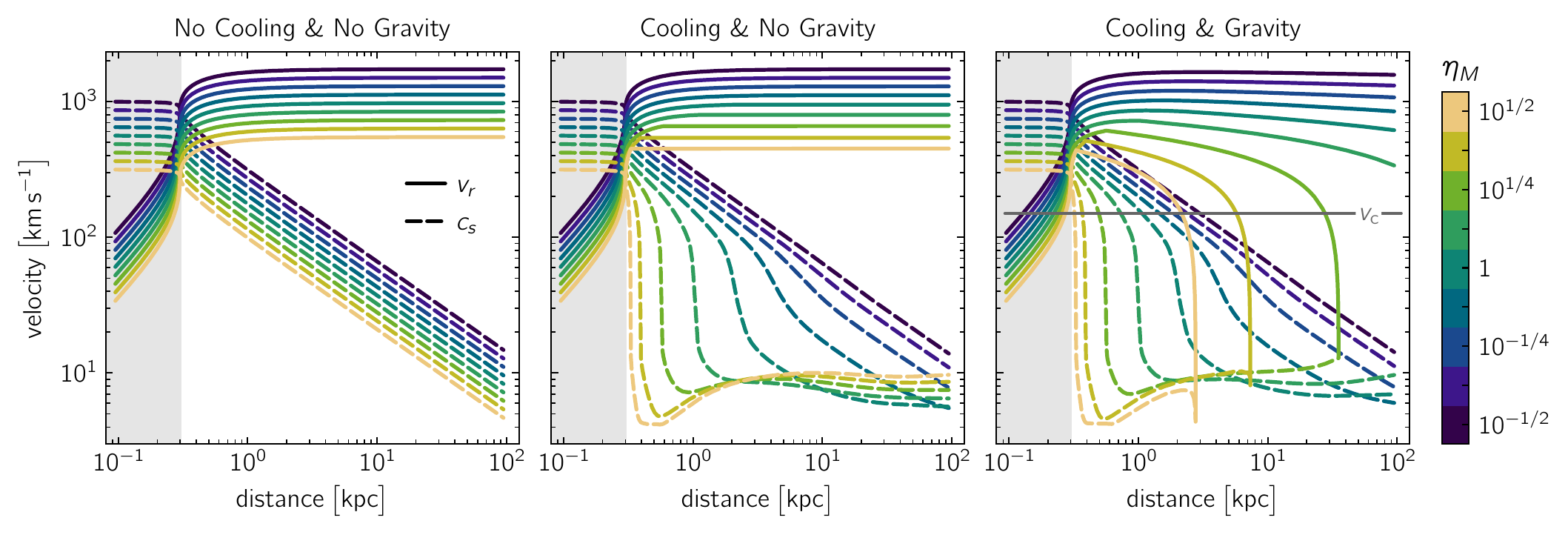} 
\caption{Velocity $v_r$ and sound speed $\cs$ profiles for galactic winds driven by a constant energy and mass injection within $\rstar = 300$ pc (gray shaded region) with $\etaE = 1$ and $\etaM$ ranging from $10^{-1/2}$ to $10^{1/2}$. The {\bf left} panel shows solutions neglecting the impact of gravity and radiative cooling, and are essentially what were presented in \citet{ChevalierClegg:1985}. We will use these solutions as a reference for our more complex solutions. How the profiles change when cooling, or cooling plus gravity (assuming an isothermal potential with a circular velocity $\vc = 150 \kms$), are included is shown in the {\bf middle} and {\bf right} panels, respectively \citep[see][]{Thompson:2016}. Note that for ease of comparison cooling and gravity are not included within the injection region $r<\rstar$. The velocity and sound speed are nearly independent of the star formation rate $\Mdotstar$ apart from a small dependence that arises from differences in the shape of the cooling curve.}
\label{fig:CC85_Cooling_and_Gravity}
\end{figure*}

The left panel of \autoref{fig:CC85_Cooling_and_Gravity} shows example galactic wind solutions for the limit where radiative cooling and gravity are negligible. In these solutions energy and mass are injected uniformly out to $\rstar = 300$ pc. The energy loading is held constant at $\etaE = 1$, while the mass loading ranges from $\etaM = 10^{-1/2}$ to $10^{1/2}$. We show the radial velocity $v_r$ and sound speed $\cs$ which do not depend on the star formation rate $\Mdotstar$ or solid angle $\Omwind$ (the density and pressure scale linearly with $\Mdotstar$ and inversely with $\Omwind$). These are essentially the same calculations as presented in \citetalias{ChevalierClegg:1985}. We will use these solutions as reference for our more complex models to come. The behavior of the solution beyond the galaxy, $r > \rstar$, is exactly as expected from \autoref{eq:simple_supersonic}. 

Radiative cooling and gravity can significantly impact the wind evolution outside of the galaxy as it expands supersonically into the surrounding medium. The middle and right panels of \autoref{fig:CC85_Cooling_and_Gravity} shows the velocity and sound speed profiles of various galactic wind solutions including only cooling (middle), and cooling and gravity (right). Cooling and gravity are not included within the injection region. We adopt solar metallicity cooling curves from \citet{Wiersma:2009} that includes heating from the $z=0$ UVB (higher redshifts would introduce small changes to the heating rate and equilibrium temperature). \citet{Thompson:2016} studied this case extensively and the solutions we find agree with their primary finding that the impact of cooling and gravity increase as the mass loading $\etaM$ increases. There is a critical mass loading $\etaM \lesssim 1$ above which the wind cools significantly relative to the adiabatic solutions. Recently, \citet{Lochhaas:2020} studied the impact of radiative cooling within the injection region, they found that the mass leaving the galaxy can be reduced by cooling, but that outside of the galaxy the wind evolves similarly albeit with a reduced mass loading. 

\subsection{Hot phase mass loss} \label{sec:hotphase_mass_loss}
We now increase the complexity of our model by considering source terms, focusing for simplicity on the region beyond the sonic radius. In particular we start with the case in which the hot phase losses mass, along with the associated momentum and energy. This is highly relevant to our goal of building a multiphase wind model since in such a model the hot phase will, in some cases, lose mass to the cold phase embedded within it. 
When there is a hot phase density sink term and there are no additional sources of momentum or energy transfer the momentum and energy source terms associated with this mass transfer are simply $\pdot = \rhodot v$ and $\edot = \rhodot \vB^2$, respectively. The evolution equations simplify to
\begin{align}
    \dloglogr{v} &= \frac{1}{1-\Mach^{-2}} \lr{ \frac{2}{\Mach^2} - \frac{\vc^2}{v^2} - \frac{\rhodot}{\rho \frac{v}{r}} \frac{1}{\Mach^2}}
\end{align}
\begin{align}
    \dloglogr{\rho} &= \frac{1}{1-\Mach^{-2}} \lr{-2 + \frac{\vc^2}{v^2} + \frac{\rhodot}{\rho \frac{v}{r}} }
\end{align}
\begin{align}
    \dloglogr{P} &= \frac{\gamma}{1-\Mach^{-2}} \lr{-2 + \frac{\vc^2}{v^2} + \frac{\rhodot}{\rho \frac{v}{r}}}.
\end{align}
In the highly supersonic and supervirial limit we get
\begin{subequations}
\begin{align}
    \dloglogr{v} &= 0 \\
    \dloglogr{\rho} &=  -2 + \frac{\rhodot}{\rho \frac{v}{r}} \\
    \dloglogr{P} &= \gamma \lr{-2 + \frac{\rhodot}{\rho \frac{v}{r}}} \\
    \dloglogr{T} &= (\gamma-1) \lr{-2 + \frac{\rhodot}{\rho \frac{v}{r}}}.
\end{align}
\end{subequations}
This highlights the fact that when $\rhodot < 0$ the density, pressure, and temperature profiles become steeper, although entropy remains flat since this transfer is by design adiabatic. 

From a mathematical perspective there is nothing preventing the $\rhodot$ from being positive, although this would mean that the source of this mass is comoving and at the same temperature as the background phase since we have adopted $\pdot = \rhodot v$ and $\edot = \rhodot \vB^2$. If $\rhodot = \rho v/r$ then $\rho \propto r^{-1}$, and $P \propto r^{-\gamma}$, which is similar to what is seen in the inner $\sim 2$ kpc in the simulations presented \citet{Schneider:2020}. This is likely caused by hot, co-moving material being added to the biconical outflow region by powerful feedback events from outside the biconical region. 

\subsection{Source terms with relative velocity} \label{sec:source_terms_relative_velocity}
We now introduce the most general form of the hot wind source terms, which account for a wide range of possible physical processes: (i) mass loss to the cold clouds and the associated momentum and energy loss, (ii) mass gain from the cold clouds and the associated momentum and energy gain, (iii) momentum transfer that is not associated with a mass transfer and the associated energy transfer, and (iv) energy transfer that is not associated with a mass or momentum transfers.
The material transferred from the cold clouds to the hot wind can (and in fact always will) have a different velocity and temperature, which has a fundamentally different impact on the structure of the wind relative to when mass is purely lost from the wind (or it is gained from material with identical velocity and temperature). For the sake of building towards our overall goal of modeling the co-evolution of cold clouds embedded in a hot wind we introduce the `cl' subscript for quantities associated with the clouds. These clouds are taken to be colder and slower than the hot wind, but nothing that follows requires that to be the case. Likewise, we call the additional momentum source term $\pdotdrag$, and the additional energy source term $\mathcal{L}$, since they will represent the drag force and radiative cooling/heating, respectively, but they could, in principle, be anything.

The general source terms are given by
\begin{subequations}
\label{eq:split_source} 
\begin{align}
    \rhodot &= \rhodotp - \rhodotm\\
    \pdot &= \rhodotp \vcl  - \rhodotm v - \pdotdrag \\
    \edot &= \rhodotp \vBcl^2 - \rhodotm \vB^2  - \pdotdrag \vcl - \mathcal{L} 
\end{align}
\end{subequations}
where each term is positive. The momentum source term is comprised of three terms: the increase of momentum from gaining mass which is moving at a velocity $\vcl$, the decrease in momentum from losing mass, and the additional momentum source term, which we take here to be due to the drag force, but could also be due to radiation pressure, for example. The energy source term is comprised of four terms: the increase of energy from gaining mass with $\vBcl^2$, the loss of energy when mass is transferred to the clouds, the loss of energy from the momentum transfer (drag force), and the additional energy source term (cooling/heating). These expressions are similar to what was used by \cite{Mathews:1971} and \cite{Cowie:1981}. The term for the energy transfer associated with the drag force can be derived using the fact that the force on the cloud is equal and opposite to the force on the wind, and that the total energy (thermal plus kinetic) must be constant. This is equivalent to saying that the drag force does work on the hot phase. 

Using these source terms the wind evolution equations become
\begin{align}
    \dloglogr{v} &= \frac{1}{1{-}\Mach^{-2}} \Bigg( \overbrace{\adibox{\frac{2}{\Mach^2}}}^{\rm adiab.} {-} \overbrace{\mombox{\frac{\vc^2}{v^2}}}^{\rm grav.} {+} \overbrace{\denbox{\frac{\rhodotm}{\rho v /r} \frac{1}{\Mach^2}}}^{\substack{ \text{cloud growth}\\ \text{wind mass loss}}} {-} \dmebox{\frac{\rhodotp}{\rho v /r}} \Bigg( \overbrace{\denbox{\frac{1}{\Mach^2}}}^{\substack{ \text{cloud shredding}\\ \text{wind mass} \\ \text{loading}}} {+} \overbrace{\mombox{\frac{v{-}\vcl}{v}}}^{\substack{\text{momentum}\\\text{transfer}}} {+} \overbrace{\enebox{\frac{\gamma{-}1}{2} \frac{\lr{v{-}\vcl}^2}{v^2}}}^{\substack{\text{kinetic energy}\\\text{thermalization}}} {-} \overbrace{\enebox{\frac{\cs^2 {-} \cscl^2}{v^2}}}^{\substack{\text{thermal}\\ \text{energy}\\\text{mixing}}} \Bigg) \nonumber \\ 
    & \hskip0.35\textwidth  {+} \underbrace{\enebox{\frac{(\gamma{-}1) \mathcal{L}}{\rho v^3 /r}}}_{\text{cooling}} - \underbrace{\enebox{\frac{(\gamma{-}1) \pdotdrag (v-\vcl) }{\rho v^3 /r}}}_{\text{drag work}} - \underbrace{\mombox{ \frac{\pdotdrag}{\rho v^2 /r}}}_\text{drag force} \Bigg) \qquad  \label{eq:velocity_gradient_split_source}\\
    \dloglogr{\rho} &= \frac{1}{1{-}\Mach^{-2}} \Bigg({-}\adibox{2} {+} \mombox{\frac{\vc^2}{v^2}} {-} \denbox{\frac{\rhodotm}{\rho v /r}} {+} \dmebox{\frac{\rhodotp}{\rho v /r}}\lr{\denbox{1} {+} \mombox{\frac{v{-}\vcl}{v}} {+} \enebox{\frac{\gamma{-}1}{2} \frac{\lr{v{-}\vcl}^2}{v^2}} {-} \enebox{\frac{\cs^2 {-} \cscl^2}{v^2}} } \nonumber \\ 
    & \hskip0.35\textwidth -  \enebox{\frac{(\gamma{-}1) \mathcal{L}}{\rho v^3 /r}} + \enebox{\frac{(\gamma{-}1) \pdotdrag (v-\vcl) }{\rho v^3 /r}} + \mombox{ \frac{\pdotdrag}{\rho v^2 /r}}\Bigg) \qquad  \label{eq:density_gradient_split_source} \\
    \dloglogr{P} &= \frac{\gamma}{1{-}\Mach^{-2}} \Bigg({-}\adibox{2} {+} \mombox{\frac{\vc^2}{v^2}} {-} \denbox{\frac{\rhodotm}{\rho v /r}} {+} \dmebox{\frac{\rhodotp}{\rho v /r}}\lr{\denbox{1} {+} \mombox{\frac{v {-} \vcl}{v}} {+} \enebox{\frac{\gamma{-}1}{2}\frac{(v{-}\vcl)^2}{\cs^2}} {-} \enebox{\frac{\cs^2 {-} \cscl^2}{\cs^2}}} \nonumber \\ 
    & \hskip0.35\textwidth -  \enebox{\frac{(\gamma{-}1) \mathcal{L}}{\rho \cs^2 v /r}} + \enebox{\frac{(\gamma{-}1) \pdotdrag (v-\vcl) }{\rho v^3 /r}} + \mombox{ \frac{\pdotdrag}{\rho v^2 /r}} \Bigg) \qquad \label{eq:pressure_gradient_split_source}
\end{align}
where we have highlighted different terms based on the their connection to the density ($\denbox{\text{green}}$), momentum ($\mombox{\text{orange}}$), or energy ($\enebox{\text{purple}}$) source terms, or if they arise regardless of source terms ($\adibox{\text{pink}}$). Each of these terms have a straight forward interpretation. The terms highlighted in pink are the adiabatic/source-free contributions. The density source term contributions, highlighted in green, account for the hot phase losing mass or gaining mass due to the growth or shredding of embedded cold clouds. The momentum source term contributions, highlighted in orange, account for the gravitational acceleration, the momentum transfer between the wind and the cloud, and the drag force. Lastly, the energy source term contributions, highlighted in purple, account for the thermalization of the kinetic energy transferred from the cold clouds to the hot phase, the thermal energy transfer as the cold clouds are mixed into the hot phase, the work done by the drag force, and the radiative cooling of the hot phase.

Using these expressions we can further write down the expression for how the entropy $K$ evolves with radius
\begin{align}
    \dloglogr{K} &= \gamma \frac{\rhodotp}{\rho v /r} \lr{ \frac{\gamma{-}1}{2} \lr{\frac{\lr{v-\vcl}^2}{\cs^2}} - \lr{\frac{\cs^2 - \cscl^2}{\cs^2}}} + (\gamma - 1) \frac{\pdotdrag (v-\vcl)}{P v / r} - \frac{r/v}{\tcoolwind} \label{eq:entropy_gradient_split_source}
\end{align}
Where we have used the usual definition of the wind cooling time $\mathcal{L} = P / ((\gamma-1) \tcoolwind)$.

\subsection{Metallicity equations}

In many interesting cases the source of mass (i.e. the clouds) may have a different metallicity than the wind. In this case the metallicity of the wind will change with radius. We can include this into the preceding machinery in a straightforward manner. The equation for metal mass conservation is
\begin{align}
    &\frac{1}{r^2} \dr{} \lr{r^2 \rhoZ v} = \rhoZdot = \rhodotp \Zcl  -  \rhodotm Z
\end{align}
where $\rhoZ$ is the metal mass density, the wind metallicity is $Z = \rhoZ / \rho$, and $\Zcl$ is the cloud metallicity. Here $\rhoZdot$ is the metal mass source term. The effective source term for the metallicity is $\Zdot \equiv \rhoZdot / \rho$. Including this conservation equation does not change the equations governing the density, velocity, and pressure. The metal mass evolution equation is 
\begin{align}
    \dloglogr{\rhoZ} &=  \dloglogr{\rho} + \frac{\rhodot}{\rho v / r} \lr{ \frac{\rhoZdot/\rhodot}{Z} - 1}= \dloglogr{\rho} + \frac{\rhodotp}{\rho v / r} \lr{\frac{\Zcl}{Z}-1}, \label{eq:metallicity_gradient}
\end{align}
which reduces to the density equation in the limit where the cloud source term has the same metallicity as the wind. 

\section{Evolution of a Cloud in a Wind} \label{sec:cloudevo}

In this section we present a simple model for the time evolution of a cold cloud moving relative to a background hot phase. In the subsequent section we will use this model as the basis for mass, momentum, and energy source terms in the wind evolution equations discussed in the previous section. The basis for our cloud-wind interaction model comes from the extensive recent work on the problem of rapidly cooling mixing layers. In particular we will draw heavily from \citet{Gronke:2018,Gronke:2020}, \citet{Fielding:2020}, and \citet{Tan:2021}.

There are two key assumptions that underpin our cold cloud model. They are (i) that the cloud is in pressure equilibrium with the surrounding hot phase and (ii) that the cloud is in thermal equilibrium with photo-heating from the UVB and/or local sources, which sets the temperature of the cloud $\Tcl$. We assume that the thermal equilibrium temperature is $\Teq = 10^4$ K, which is commensurate with estimates for most galactic and circumgalactic environments. These assumptions break in interesting regimes. The speed at which pressure is communicated through a cloud is given by $\chi^{-1/2} \, \cshot$, where $\chi = \rhocl / \rho$. We, therefore, expect the cloud to remain in pressure equilibrium if the time it takes the cloud to enter into a region with a new pressure is longer than the time it takes that pressure change to be communicated throughout the cloud. If $H_P$ is the characteristic pressure scale height then we expect clouds to remain in pressure equilibrium if $\rcl < H_P \chi^{1/2} (\vcl/\cshot)$. The assumption of pressure equilibrium is increasingly suspect for larger clouds moving supersonically through a medium with a small pressure scale height. The assumption of thermal equilibrium implicitly assumes that the the radiative heating and cooling times are very short relative to all other time scales. This assumption is weakest for low pressure clouds ($P \lesssim 10$ K cm$^{-3}$) in which these radiative timescales become comparable to the flow times. In our baseline model we assume that both pressure and thermal equilibrium are maintained instantaneously. 
These assumptions allow us to calculate the density of a cloud $\rhocl = \mump P / \kb \Tcl$.

\subsection{General Single Cloud Framework}\label{sec:single_cloud}

The fundamental quantity in our cloud-wind interaction model is the mass transfer. This mass transfer arises from the competition between processes that cause the cold cloud to grow or lose mass. In this subsection we write down the general framework for cloud evolution for an arbitrary set of mass transfer rates. We separate the mass growth and loss terms and write the total cloud mass transfer rate to be
\begin{align}
    \Mdotcl = \Mdotclgrow - \Mdotclloss \label{eq:Mdotcl_general}
\end{align} 
where $\Mdotclgrow,\Mdotclloss>0$. Conceptually, the cloud losses mass because it is shredded as the hot wind moves past it. This relative motion promotes the Kelvin Helmholtz instability (KHI), which in turn leads to the development of a turbulent mixing layer. The turbulence in these mixing layers acts as diffusivity which eats away at the cloud. However, the turbulent mixing populates the intermediate temperature regimes, which cool much more rapidly than the cloud or the background wind. Cooling in these \emph{turbulent radiative mixing layers} (TRMLs) offsets the mass loss and in some cases can lead to a net growth of the cloud. In the following subsection we lay out the exact parameterizations of the mass transfer rates that we will adopt, while in this subsection we keep the formulation as general as possible.

The growth of the cloud arises from a transfer from the hot wind phase to the cold cloud. As this transfer proceeds the mass flowing into the cloud brings with it the momentum from the hot phase by adding material moving at the hot phase velocity $\vecv$. Likewise, as the cloud loses mass it loses momentum by shedding material moving at the cloud velocity $\vecvcl$. The momentum transfer\footnote{We adopt the symbol $\varphi$ for momentum, which has units of mass times velocity, as opposed to the momentum density that is denoted by $p$ and has units of density times velocity.} and thus the acceleration associated with mass transfer are given by
\begin{equation}
    \vecpdotcltrans = \vecv \Mdotclgrow - \vecvcl \Mdotclloss \quad \Rightarrow \quad \vdotcltrans = \frac{\vecpdotcltrans - \vecvcl \Mdotcl}{\Mcl} = (\vecv-\vecvcl) \frac{\Mdotclgrow}{\Mcl}.
\end{equation}
The cloud will also experience a drag (or ram pressure) force from the wind given by
\begin{align}
    \vecpdotcldrag = \frac{1}{2} \Cdrag \rho |\vecv-\vecvcl|^2 \frac{\vecv-\vecvcl}{|\vecv-\vecvcl|} \Across
\end{align}
where $\Across = \pi \rcl^2$ is the cross sectional area of the cloud. We assume a fiducial value of $\Cdrag = 1/2$ in our numerical calculations below, which is broadly consistent with predictions of high Reynolds number flows.
The cloud also experiences an acceleration due to gravity, which is simply 
\begin{align}
    \vecpdotclgrav = -\Mcl \vec{\nabla}\Phi = -\Mcl \frac{\vc^2}{r} \hat{r}.
\end{align}
Where, in the second equality, we have written the gradient of the gravitational potential using $\vc^2 \equiv g\,r$, where $g$ is the gravitational acceleration, which is a convenient form for the spherically symmetric external potentials we will use later.

The total force is given by the sum of $\vecpdotcltrans$, $\vecpdotcldrag$, and $\vecpdotclgrav$. In our model we adopt the simplifying assumption that the flow is one dimensional, which allows us to write the force along this direction ($\hat{r}$ for spherical coordinates or $\hat{x}$ for plane parallel coordinates) as
\begin{align}
    \pdotcl = v \Mdotclgrow - \vcl \Mdotclloss + \frac{1}{2} \Cdrag \rho (v-\vcl)^2 \Across - \Mcl \frac{\vc^2}{r}.
\end{align}
The acceleration in this case is 
\begin{align}
    \vdotcl = (v-\vcl) \frac{\Mdotclgrow}{\Mcl} + \frac{3 \Cdrag}{8}\frac{(v-\vcl)^2}{\chi \rcl} - \frac{\vc^2}{r} \label{eq:vdotcl_general}
\end{align}
where we have assumed a spherical cloud with $\Mcl = (4 \pi /3) \rhocl \rcl^3$. 

Along with the mass and momentum transfer we must account for the energy transfer. From the perspective of the cloud the mass and momentum transfer rates along with the assumptions of pressure and thermal equilibrium obviate the need for explicitly tracking the energy transfer rates. However, from the perspective of the wind the energy transfer is important. The cloud sheds and gains thermal and kinetic energy as it loses and gains mass. Some of the gained kinetic energy will be thermalized, but this is immediately radiated away thereby keeping the cloud at $\Teq$. Likewise, some of the kinetic energy that the cloud loses will be thermalized in the wind, however this energy is not assumed to radiate away and can provide appreciable heating (see \autoref{eq:entropy_gradient_split_source}). Similarly, the change in momentum from the drag force will also result in an energy transfer between the hot wind and the cold clouds. 
The total energy\footnote{We adopt the symbol $\mathcal{E}$ for energy, which includes kinetic and thermal energy and has units of mass times velocity squared, as opposed to the energy density that is denoted by $\varepsilon$ and has units of density times velocity squared} transfer rate from the cloud to the hot phase of the wind is
\begin{align}
    \edothot = -\vB^2 \Mdotclgrow + \vBcl^2 \Mdotclloss - \vecvcl \cdot \vecpdotcldrag. \label{eq:epdothot_general}
\end{align}
The Bernoulli velocity $\vB$ (kinetic energy plus enthalpy) appears in this expression (as opposed to the kinetic energy plus thermal energy) in order to account for the $P dV$ work. Since we are assuming constant pressure this is akin to isobaric cooling. Therefore, it is the enthalphy, rather than just the internal energy, that is transferred to or from the cloud during the mass transfer. 

In practice, the hot phase energy transfer rate $\edothot$ is all that is needed, however it is instructive to understand how much energy is lost due to cooling. Unlike with the mass and momentum transfer, the energy transfer of the cloud is not equal and opposite to that of the hot wind because some of the energy is radiated away. There are two sources of energy that are radiated away: the excess enthalpy, and the relative kinetic energy. The enthalpy that is advected into the cold cloud from the hot phase along with the mass growth is almost entirely radiated away, which is why there is any mass growth in the first place. In the TRML, material cools down from the hot phase temperature to the thermal equilibrium temperature, which is, by assumption, the cloud temperature. Additionally, kinetic energy is advected into the cold cloud as it grows. The relative kinetic energy between the phases must also be radiated away to maintain the correct cloud energy. The amount of energy that is radiated away is 
\begin{align}
    \edotcool = \dot{\mathcal{E}}_{\rm cool}^{\rm (enthalpy)} + \dot{\mathcal{E}}_{\rm cool}^{\rm (kinetic)} = \lr{\frac{\cshot^2 - \cscl^2}{\gamma - 1} +\frac{1}{2} \vrel^2}  \Mdotclgrow \label{eq:epdotcool_general}
\end{align}
The change in the total energy of the cloud is thus
\begin{align}
    \edotcl = \vB^2 \Mdotclgrow - \vBcl^2 \Mdotclloss - \edotcool + \vecvcl \cdot \vecpdotcldrag = -\edothot - \edotcool. \label{eq:epdotcl_general}
\end{align}

In the case where the cloud and wind do not have the same metallicity there is a transfer of metals. The total metal mass of a cloud is $\MZcl \equiv \Zcl \Mcl$. The metal mass transfer rate and metallicity change rate are 
\begin{align}
    \MZdotcl = Z \Mdotclgrow + \Zcl \Mdotclloss \quad \Rightarrow \quad \Zdotcl = \frac{\MZdotcl - \Zcl \Mdotcl}{\Mcl} = \lr{Z-\Zcl}\frac{\Mdotclgrow}{\Mcl}. \label{eq:Zdotcl_general}
\end{align}
The evolution of the cloud metallicity can be a useful tracer of the amount of mixing that has occurred.

\subsection{Mass Transfer Model}\label{sec:mass_transfer_model}

With the general framework in hand we can now move onto specific parameterizations of the mass fluxes. Robust numerical simulations have not yet led to a clear picture of what the exact parameterizations for the mass fluxes should be for a specific set of cloud-wind parameters. We, therefore, introduce the mass fluxes in a form that highlights specific choices that need to be made and can, hopefully, be constrained by simulations in the future. 

\subsubsection{Cloud Growth Rate Model}
We start with a cloud growth rate model based on \citetalias{Fielding:2020}, \citetalias{Tan:2021}, and \citet{Gronke:2020}
\begin{align}
    \Mdotclgrow = \rho \Acool \vin \;\; {\rm with} \;\; \vin = \vturb \xi^{\alpha} \;\; {\rm where} \;\; \alpha = \begin{cases}
    1/4 & \text{if } \xi \geq 1 \\
    1/2 & \text{if } \xi < 1
    \end{cases} \;\; {\rm with} \;\; \xi = \frac{\rcl}{\vturb \tcl}.
\end{align}
Here $\vin$ is the inflow velocity from the hot phase onto the cloud, $\tcl$ is the characteristic cooling time of gas in mixing layer, $\vturb$ is the turbulent velocity in the mixing layer, $\xi$ is the crucial parameter that encodes the relative strengths of turbulent mixing and radiative cooling, and $\Acool$ is the cooling area of the cloud. 
The change of $\alpha$ from 1/2 in the slow cooling regime ($\xi <1$) to 1/4 in the rapid cooling regime ($\xi>1$) is one of the major findings of recent simulation work on this topic (\citetalias{Fielding:2020}; \citealt{Gronke:2020}; \citetalias{Tan:2021}).
What sets $\tcl$ has yet to be conclusively determined \citep{Abruzzo:2021}, so for simplicity sake we follow \citetalias{Gronke:2018} and define the characteristic cooling time to be the cooling time of material with a temperature and metallicity equal to the geometric mean of the wind and cloud,  $\tcl = \tcool(T = \sqrt{T \Tcl}, Z = \sqrt{Z \Zcl})$. An appealing alternative choice is to define $\tcl$ to be the minimum cooling time at the current pressure for a metallicity equal to that of the cloud. Different choices for $\tcl$ will make minor quantitative changes rather than introduce a qualitative difference. The one exception would be if the growth rate depended on the cooling time of the hot gas as has been suggested by \citet{Li:2020} and  \citet{Sparre:2020}. We will, however, not explore this option.

Cloud crushing simulations in which cooling is strong clearly demonstrate that the wake of the clouds extends out for many $\rcl$ behind the clouds, and that this is where the cooling and mass growth happen. The amount of this elongation increases with density contrast $\chi$. This allows us to write the mass growth as 
\begin{align}
    \Mdotclgrow = \frac{\Mcl \vrel}{\chi^{1/2} \rcl} \lr{\frac{\vturb}{\vrel}} \lr{\frac{\Acool}{\frac{4}{3} \pi \rcl^2 \chi^{1/2}}} \xi^\alpha
\end{align}
where the cloud mass is defined $\Mcl = (4 \pi/3) \rhocl \rcl^3$.
In order to use this equation we must specify how the intrinsic mixing layer turbulent velocity relates to the relative velocity $(\vturb/\vrel)$ and how the cooling area relates to the cloud area $(\Acool / \rcl^2)$. Motivated by \citetalias{Fielding:2020} and \citetalias{Tan:2021} we make the fiducial assumption that 
\begin{align}
    \fturb = \frac{\vturb}{\vrel} = \text{constant} \approx 0.1.
\end{align}
How elongated the clouds are is less clear from simulations since as the clouds are being stretched out they are also being accelerated and the relative velocity is dropping. We therefore assume that the characteristic timescale over which the wake forms is given by the cloud crushing time $\tcc$, which is equivalent to the KHI growth timescale. This means that the cooling area is 
\begin{align}
    \Acool \propto \rcl \vrel \tcc \text{ or } \Acool =  \fcool 4 \pi \rcl^2 \chi^{1/2}.
\end{align}
For simplicity we assume a constant $\fcool = 1/3$. For comparison, \citetalias{Gronke:2020} argued that the cooling area reaches $\rcl^2 \chi$, but only when the cloud is fully entrained. We can now combine everything into one final expression for the growth rate
\begin{align}
    \Mdotclgrow = 3 \fturb \fcool \lr{\frac{\Mcl \vrel}{\chi^{1/2} \rcl}}  \xi^\alpha = 3 \fturb \fcool \lr{\frac{\Mcl \vrel}{\chi^{1/2} \rcl}} \lr{\frac{\rcl}{\fturb \vrel \tcl}}^\alpha
\end{align}

\subsubsection{Cloud Mass Loss Rate Model}
We now turn our attention to the cloud mass loss rate. The general form for the mass loss rate is similar to that of the mass growth rate and is given by
\begin{align}
    \Mdotclloss = \rhocl \; \fmix 4 \pi \rcl^2 \; \vturbcl
\end{align}
where we have assumed the area over which the cloud is mixed is given by $\fmix 4 \pi \rcl^2$, we take $\fmix = \fcool = 1/3$, and $\vturbcl$ is the turbulent velocity responsible for mixing the cloud with the background medium. This $\vturbcl$ can be different from the turbulent velocity in the mixing layer $\vturb$ that is responsible for mixing hot gas into the cold gas. Motivated by \citetalias{Fielding:2020} we assume that turbulent kinetic energy density in hot and cold phases is constant which implies $\rho \vturb^2 = \rhocl \vturbcl^2$. This assumption is equivalent to the arguments made by \cite{Begelman:1990} and \cite{Gronke:2018} to get that the characteristic values of the mixing layer properties is equal to the geometric mean of the cloud and hot phase properties (i.e. $T_{\rm mix} = \sqrt{\Tcl T}$). Combining this all together gives 
\begin{align}
    \Mdotclloss = 3 \fturb \fmix \lr{\frac{\Mcl \vrel}{\chi^{1/2} \rcl}} \approx \frac{\Mcl}{\tcc},
\end{align}
which is similar to the canonical cloud crushing mass loss rates \citep{Klein:1994}.

\subsubsection{Full Cloud Mass Transfer Model}
With the cloud mass growth and loss expressions in hand we can now compactly express the total mass growth
\begin{align}
    \Mdotcl = 3 \fturb \fcool \lr{\frac{\Mcl\vrel}{\chi^{1/2} \rcl}} \lr{\xi^\alpha -1} = \Mdo \lr{\xi^\alpha -1}.
\end{align}
Where we have for convenience introduced $\Mdo = (3 \fturb \fcool \Mcl \vrel) / (\chi^{1/2} \rcl)$, so $\Mdotclgrow = \Mdo \xi^\alpha$, and $\Mdotclloss = \Mdo$. 
This expression has the nice behavior that clouds with $\xi > 1$ grow, while clouds with $\xi < 1$ shrink, which is seen in radiative cloud crushing simulations \citep[\eg][]{Abruzzo:2021}.\footnote{In our model mass is transferred to and from the cloud simultaneously. Thus when $\xi > 1$ and the cloud is net growing there is still a transfer of mass, momentum, and energy to the hot wind from the cloud. An alternative choice to our `two-way' transfer model is a `one-way' transfer model in which transfer from the cloud to the hot phase only occurs when the net mass transfer rate is negative, i.e. $\xi < 1$. Further numerical experimentation is necessary to determine which model provides the best description, but in our testing we found only minor quantitative differences when adopting these two options, and so we stick with the two-way transfer model throughout.}

\begin{figure*}
\centering
\includegraphics[width=\textwidth]{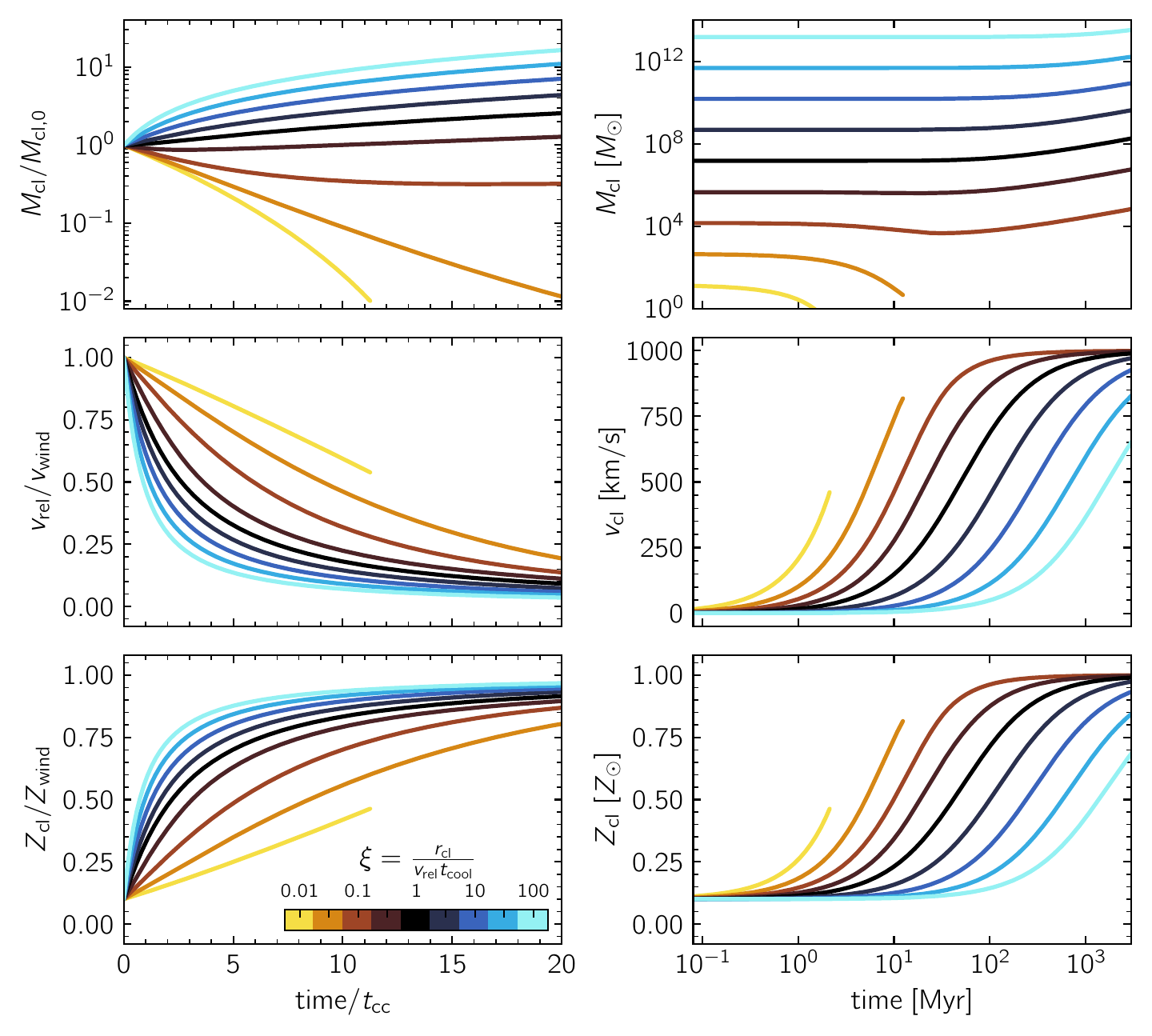} 
\caption{Example time evolution of the cloud mass, velocity, and metallicity for cold clouds with a density contrast $\chi = 100$, an initial relative velocity of 1000 km/s, and a range of initial masses, which corresponds to a range of initial $\xi$ values from 0.01 to 100. The clouds have an initial metallicity of 0.1 $Z_\odot$, while the wind has a metallicity of $Z_\odot$. The pressure of the clouds and wind is $P/k_B = 10^3$ K cm$^{-3}$, the cloud temperature is $10^4$K, and the hot wind temperature is $10^6$K. The left panels show the evolution normalized by the relevant quantities, while the right panels show the evolution in physical units. The normalized evolution depends only on $\xi$ with only minor differences arising due to changes in the cooling curve at different pressures. The lines end if the cloud has lost 99 \% of its initial mass. These evolutionary tracks agree well, on a qualitative level, with those shown for full three dimensional radiative cloud crushing simulations \citep[\eg][]{Gronke:2018, Abruzzo:2021}.}
\label{fig:fiducial_cloud_wind}
\end{figure*}

\autoref{fig:fiducial_cloud_wind} shows a series of example cloud mass, velocity, and metallicity evolutionary sequences for a range of $\xi$ values for a cloud moving through an infinite, uniform, hot background medium. In these examples the cloud has no back-reaction on the hot medium. There is no gravitational acceleration included in these examples. The density contrast is $\chi = 100$ and the initial relative velocity is $\vrel = 1000$ km/s. The pressure is $P/k_B = 10^3$ K cm$^{-3}$, $T_{\rm hot} = 10^6$ K, and $\Tcl = 10^4$ K. The left column shows the normalized quantities and the right column shows the quantities in physical units. The normalized evolution depends only on $\xi$. Large clouds with initial $\xi\gg1$ grow significantly, become very well entrained, and gain significant metallicity in the first few $\tcc$. Small clouds on the other hand with initial $\xi \ll 1$ are quickly shredded and destroyed before becoming entrained. Intermediate clouds with $\xi \lesssim 1$ initially lose mass, but as the cloud becomes more entrained the instantaneous $\xi$ increases until the cloud begins to grow. In physical units, large clouds ($\xi > 1$) evolve on a much longer timescale than small clouds ($\xi < 1$), which, as we will discuss below, means that they can impact the hot wind evolution far from the galaxy. Lastly, there is a close correspondence between the relative velocity $\vrel$ and the cloud metallicity $\Zcl$, which reflects the fact that the acceleration is dominated by the mass transfer associated with cooling in the TRML since this changes the metallicity, as opposed to ram pressure, which does not lead to a change in metallicity.

There is a subtle but noteworthy difference between our model and the rough picture put forward in \citetalias{Gronke:2018}, which is that our models grow when $\xi > 1$ whereas they assume clouds grow when $\tcc / \tcl > 1$. Our two criteria differ by a factor of $\fturb \; \chi^{-0.5}$. When $\fturb = 0.1$ and $\chi = 100$ the two criteria are the same in practice. However, when keeping $\fturb = 0.1$ (as is motivated by Fig. 12 in \citetalias{Tan:2021}), and increasing $\chi>100$ a cloud with $\tcc = \tcl$ will have $\xi < 1$. Therefore in our model this cloud would \emph{initially} lose mass. However, clouds with $\xi \lesssim 1$ often end up gaining mass at late times since as $\vrel \rightarrow 0$, $\xi$ increases. It is possible therefore that the early time ($t \lesssim 5-10 \tcc$) decrease in cloud mass seen in nearly all $\chi > 100$ simulations with $\tcc \geq \tcl$ is consistent with our model since in many of these cases $\xi<1$ \citep[\eg][]{Gronke:2018, Gronke:2020, Li:2020, Sparre:2020, Kanjilal:2021, Abruzzo:2021}. This effect becomes particularly pronounced in $\chi \gtrsim 10^4$, which is common at the base of strong galactic winds.

\section{Multiphase Galactic Wind Model} \label{sec:coevolution}
Now that we have laid out and explored both the general form of the wind equations with mass, energy, and momentum source terms (\autoref{sec:volumefilling}), and the general form of the cloud-wind interaction (\autoref{sec:cloudevo}) we are ready to combine the two. To do so, we need to cast the mass, energy, and momentum fluxes that arise as a result of the cloud-wind interaction into source terms for the volume filling phase of the wind. At the same time we must account for the radial evolution of the cloud number density. Our end goal is to write down a system of equations that can be solved to describe the steady state structure of a multiphase galactic wind that specifically accounts for the cloud-wind transfers that occur as a result of cooling in the turbulent radiative mixing layers that separate the phases, drag forces, and cooling in the volume-filling hot phase. 

The solution to these system of equations quickly become complex and hard to interpret, so in this paper we restrict ourselves to an exploration of the simplest model in which there is a single population of identical clouds at each radius. This simple model captures the vast majority of the important behavior. In a follow-up work we extend this to include multiple populations of clouds, and a full distribution of clouds (Anthony Chow et al. in prep.). 

The amount of mass, momentum, and energy transferred between the cold, embedded clouds and the hot, volume-filling wind depends on the transfer rates of an individual cloud and the volumetric number density of clouds $\ncl$. The number density of clouds depends on the number flux of clouds $\Ncldot$, which in steady state is independent of time. Conceptually, the simplest choice for $\Ncldot$ is a step function, which corresponds physically to a scenario in which at small radii there are no clouds, at some radius the clouds are introduced, and from then on clouds are neither created or destroyed. We adopt a slightly more physically realistic choice that allows for an increase in $\Ncldot$ at small radii, which represents the sweeping up of an increasing number of cold clouds as the hot wind moves through the galaxy. Further modifications to $\Ncldot$ can account for clouds being created, destroyed, merged, or split while moving out in the wind, however, for the sake of ease of interpretation, we leave the exploration of these variations to a future work. 

For a given $\Ncldot$ the number density of clouds at a radius is 
\begin{align}
    \ncl = \frac{\Ncldot}{\Omwind r^2 \vcl}. \label{eq:cloud_density}
\end{align}
With the number density of clouds and the model for the cloud-wind transfer rates we can now write down the source terms that the volume-filling component of the wind experience
\begin{align}
    &\rhodot = - \Mdotcl \ncl = - \ncl \lr{\Mdotclgrow - \Mdotclloss} =  - \rhodotm + \rhodotp  \\
    &\pdot   = - \pdotcl \ncl = - \ncl \lr{v \Mdotclgrow - \vcl \Mdotclloss + \pdotcldrag} = - v \rhodotm + \vcl \rhodotp - \pdotdrag\\
    &\edot   =  \edothot \ncl = - \ncl \lr{\vB^2 \Mdotclgrow - \vBcl^2 \Mdotclloss + \pdotcldrag \vcl} = -\vB^2 \rhodotm + \vBcl^2 \rhodotp - \pdotdrag \vcl \\
    &\rhoZdot = - \MZdotcl \ncl = -\ncl \lr{ Z \Mdotclgrow - \Zcl \Mdotclloss} = -\rhodotm Z + \rhodotp \Zcl 
\end{align}
where $\Mdotcl$, $\pdotcl$, and $\edothot$ are defined using the definitions in the previous section (\autoref{eq:Mdotcl_general}, \autoref{eq:vdotcl_general}, \autoref{eq:epdothot_general}), and the different terms in each of the source terms can be associated with the general formulation in \autoref{sec:source_terms_relative_velocity}. The interaction of the cold clouds with the hot, volume-filling phase of the wind will lead the cloud mass, velocity, and metallicity to evolve with radius as
\begin{subequations}
\label{eq:cloud_evolution}
\begin{align}
    \dr{\Mcl} &= \frac{\Mdotcl}{\vcl} \\ 
    \dr{\vcl} &= \frac{\vdotcl}{\vcl} \\
    \dr{\Zcl} &= \frac{\Zdotcl}{\vcl} .
\end{align}
\end{subequations}
Lastly, the total mass flux in cold clouds at a given radius is simply $\Mdotcold = \Mcl(r) \Ncldot$.

We can now use these source terms in the equations for the steady state radial structure of velocity, density, and pressure that we defined in \autoref{eq:velocity_gradient_split_source},  \autoref{eq:density_gradient_split_source}, and \autoref{eq:pressure_gradient_split_source}. We, therefore, now have everything we need to co-evolve the volume-filling hot phase of the wind and the cold clouds embedded within it.

\subsection{Low mass loading examples} \label{sec:low_eta}

\begin{figure*}
\centering
\includegraphics[width=\textwidth]{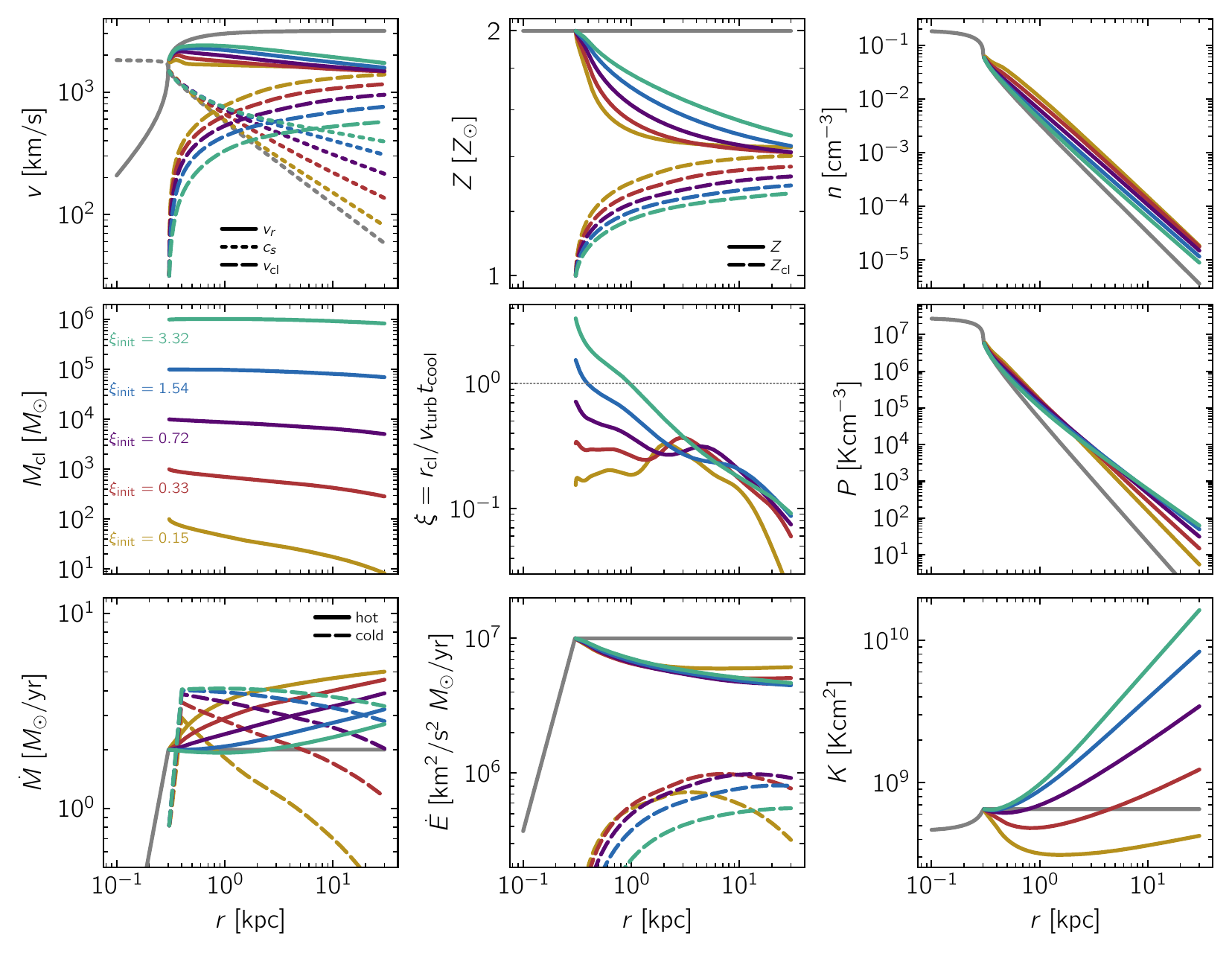} 
\caption{Radial profiles of multiphase wind solutions. The hot phase is powered by a uniform mass and energy injection within $\rstar = 300$pc that corresponds to a SFR = $20 \Msunyr$, $\etaM=0.1$, and $\etaE=1$. There is an isothermal gravitational potential with $\vc = 150$ km/s.  The gray lines show the single-phase adiabatic solution and the colored lines show the solutions when cold clouds are present. Cold clouds are introduced into the wind with an initial velocity of 30 km/s. The initial mass of the clouds in the different solutions ranges from $10^2 \Msun$, shown in gold, to $10^6 \Msun$, shown in cyan. The number flux of clouds $\Ncldot$ is set such that the initial cold phase mass flux corresponds to $\etaMcold = 0.2$. The clouds are introduced between 300 and 400 pc to avoid over mass-loading the wind and causing it to fail. This is achieved by specifying $\Ncldot$ to increase with radius. The solutions beyond several $\rstar$ are insensitive to these choices. The different evolution of the solutions highlights the importance of the cloud mass on the overall evolution of both the clouds and the hot, volume-filling component of the wind. The clouds in solutions with lower initial cloud masses are shredded and accelerated rapidly leading to a slower and colder volume-filling phase relative to the solutions with more massive initial cloud masses that exhibit slower shredding and acceleration.}
\label{fig:Case_I_low_eta}
\end{figure*}

\autoref{fig:Case_I_low_eta} shows five example multiphase wind solutions. The hot wind properties are initially the same in all of the solutions. The injection rates that dictate the initial hot wind properties correspond to a SFR = $20 \Msunyr$, $\etaE = 1$, and $\etaM = 0.1$, which are added uniformly within the injection radius $\rstar = 300$ pc. These parameters were chosen to roughly match the properties of the M82 outflow \citep{Strickland:2009}, and the high SFR part of the simulations presented by \citet{Schneider:2020}. Physically, these properties correspond to a case in which the hot outflow loses no energy to cooling in the injection region and its mass is set entirely by SN ejecta ($\Mejecta \approx 10 \ \Msun$ per SN), i.e. it has not swept up any additional ISM material yet. This is in line with findings of high resolution, multi-physics ISM patch simulations that generically find order unity energy loading factors $\etaE$ and hot phase mass loading factors on the order of $\etaM = \Mejecta / \mstar = 0.1$ \citep[\eg][]{Li:2020a, Kim:2020a}. The gray lines in all of the panels show the solution for a single-phase adiabatic wind. The colored lines show the solution once cold clouds have been introduced into the wind. The cold clouds in all cases are added gradually between 300 and 400 pc, with an initial $\etaMcold = 0.2 = 2 \etaM$ and $\vcl = 30 \;\kms$. The different color lines correspond to different initial cloud masses, ranging from $\Mcl = 10^2 \, \Msun$ (gold) to $\Mcl = 10^6 \, \Msun$ (cyan). The initial metallicity of the hot phase and the clouds are $2 Z_\odot$ and $Z_\odot$, respectively.

The introduction of cold clouds into the hot wind, shown in \autoref{fig:Case_I_low_eta}, has a significant and varied impact on the properties of both the clouds themselves and the hot wind. The transfer of material between the clouds and the wind, along with the drag force, causes the hot volume-filling phase to decelerate and the clouds to accelerate, which can be seen in the top left panel of \autoref{fig:Case_I_low_eta} in the solid and dashed lines, respectively. The rate of cloud-wind mass transfer is predominately controlled by $\xi$. As shown in the middle panel, the two most massive clouds ($\Mcl = 10^5$ and $10^6 \ \Msun$) have $\xi > 1$ when they are first introduced, whereas the less massive clouds have $\xi < 1$. As the clouds move out in the wind $\xi = \rcl / \vturb \tcl$ changes due to four competing effects: (i) the cloud radius increases as the confining pressure drops $\rcl \propto P^{-1/3}$, (ii) the cloud radius changes as the cloud mass changes $\rcl \propto \Mcl^{1/3}$, (iii) the turbulent velocity in the TRML decreases as the cloud becomes entrained and $\vrel$ decreases, and (iv) the cooling time in the TRML $\tcl$ changes as hot wind properties change. For the more massive clouds $\xi$ rapidly drops below one due to the strong increase in $\tcl$. The change in $\xi$ for the less massive clouds is more complex, but $\xi$ remains less than one throughout the clouds' evolutions. Generically, when the hot phase density is low, either due to a low value of $\etaM$ or low SFR, the increase in $\tcl$ due to the decrease of the hot phase density and pressure dominates at large radius, ensuring $\xi < 1$ for $r \gg \rstar$.

These changes in $\xi$ for different choices of initial cloud masses are reflected in the radial evolution of the cloud mass shown in the middle right panel of \autoref{fig:Case_I_low_eta}. The massive clouds undergo a short period of mass growth while their $\xi > 1$ then steadily lose mass to the hot phase at a relatively slow rate. The smaller clouds lose mass from the moment they are introduced, however, even though $\xi \sim 0.2$ for most of their evolution, by the end of the integration the smallest clouds (that started with a mass of $10^2 \ \Msun$) have retained 10\% of their initial mass, and the largest clouds retained $\sim 90$\% of their initial mass. This highlights the essential point that even when cooling is insufficient to lead to cloud growth it still plays a major role by significantly reducing the rate at which the clouds are destroyed. 

As the clouds lose and gain mass in the various solutions shown in \autoref{fig:Case_I_low_eta}, material is transferred to and from the hot phase. A readily apparent effect of this is the change in the mass fluxes of the cold and hot phases, which is shown in the lower left panel. In all cases the hot phase mass flux increases at the expense of the cold phase, while the total mass flux stays constant as required by the steady state equations. In the solution with $10^2 \Msun$ clouds the total cold phase mass flux drops rapidly, leading to a $\sim 2.5 \times$ increase in the hot phase mass flux. On the other hand, in the solution with $10^6 \Msun$ the hot phase mass flux initially decreases while the cold clouds grow, until at large radii when the hot phase mass flux increases somewhat. The energy fluxes, shown in the middle panel of the bottom row show different behavior. The hot phase loses energy even as it gains mass. This is because of the significant amount of energy that is radiated away in the TRMLs that govern much of the cloud-wind interaction. Radiative cooling in the volume filling component itself is negligible for the low hot phase mass loading of these solutions, but can become important as will be shown below. By the time the wind has reached 30 kpc the TRML cooling has drained $\sim 50$ \% of the original energy of the hot phase. The energy of the cold clouds increases as they are accelerated but can be diminished as they lose mass. 

The interaction of the cold clouds with the hot wind lead to manifold changes of the hot phase's bulk properties. Some of these changes are straightforward to understand, such as the change in the metallicity as cloud material is mixed into the hot phase. However the changes to the density and pressure result from the complex interplay of numerous competing effects. The net result is an increase in the density and pressure relative to the adiabatic, single-phase solution as shown in the upper and middle panels of the right column. In the solutions with more massive clouds, the entropy of the hot phase increases, whereas solutions with smaller clouds exhibit an entropy decrease. These changes correspond to a mild flattening of the sound speed $\cs$ (and $T$) profiles (upper left panel) relative to the adiabatic solutions ($\cs \propto r^{-2/3}$) when the cold clouds are small, and a strong flattening when the clouds are large ($\cs \propto r^{-1/3}$). 

\begin{figure*}
\centering
\includegraphics[width=\textwidth]{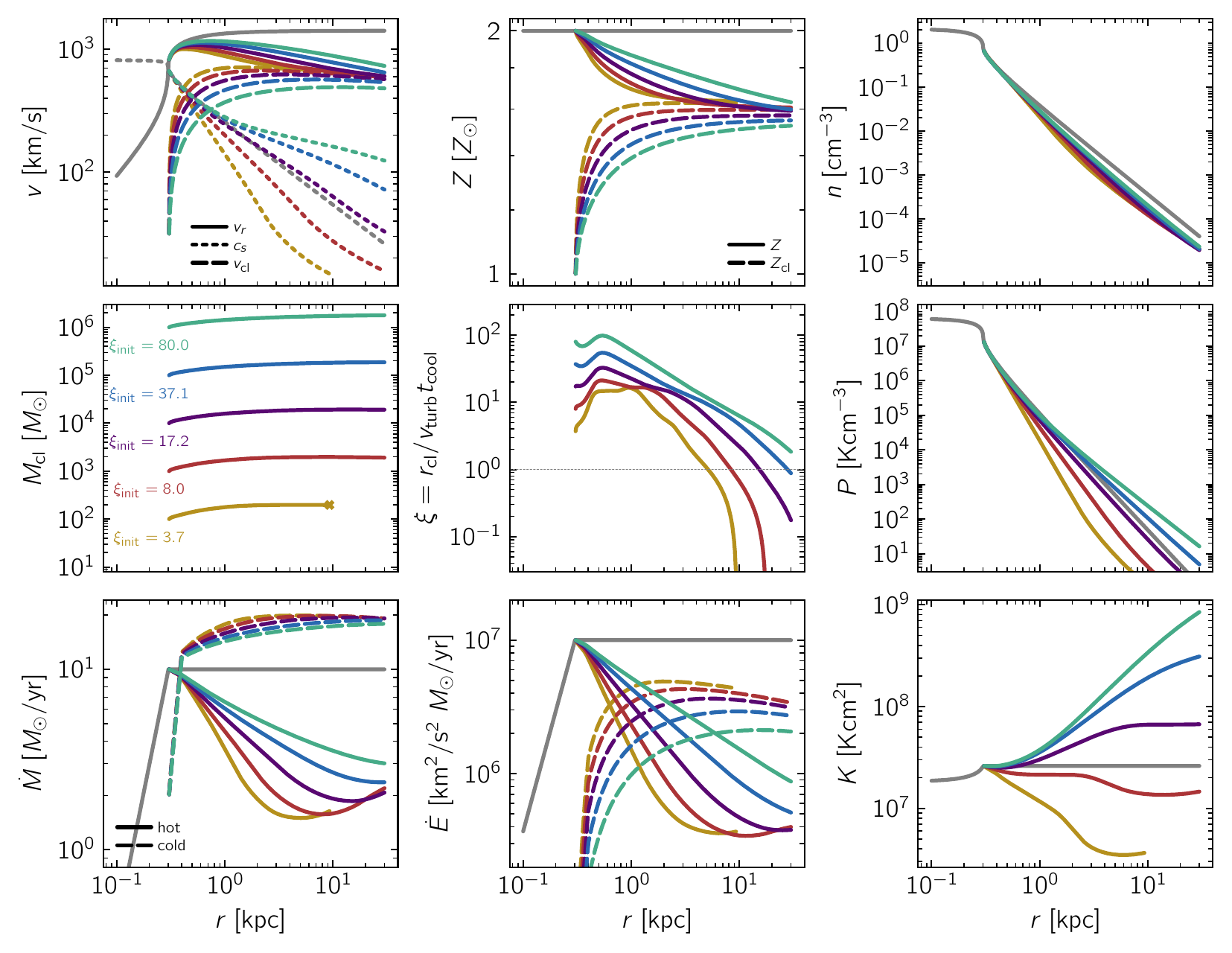} 
\caption{Radial profiles of multiphase wind solutions similar to \autoref{fig:Case_I_low_eta} but with higher mass loading factors. The hot phase is powered by a uniform mass and energy injection within $\rstar = 300$pc that corresponds to a SFR = $20 \Msunyr$, $\etaM=0.5$, and $\etaE=1$. There is an isothermal gravitational potential with $\vc = 150$ km/s. The gray lines show the single-phase adiabatic solution and the colored lines show the solutions when cold clouds are present. Cold clouds are introduced into the wind with an initial velocity of 30 km/s. The initial mass of the clouds in the different solutions ranges from $10^2 \Msun$, shown in yellow, to $10^6 \Msun$, shown in cyan. The number flux of clouds $\Ncldot$ is set such that the initial cold phase mass flux corresponds to $\etaMcold = 0.5$. The hot component is slower and denser than in the solutions shown in \autoref{fig:Case_I_low_eta}. The higher hot phase density causes there to be increase cooling in both the mixing layer and the hot phase itself. The lower velocities make the gravitational deceleration non-negligible. The solutions are cut off and marked by an `x' if the wind temperature falls to the cloud temperature. }
\label{fig:Case_I_high_eta}
\end{figure*}

\subsection{High mass loading examples} \label{sec:high_eta}

In the previous examples, gravity and cooling in the volume-filling hot phase were unimportant, however, with only a modest increase in the hot phase mass loading this is no longer true. \autoref{fig:Case_I_high_eta} shows example solutions similar to the previous set with the only differences being an initial hot phase mass loading of $\etaM = 0.5$ and cold phase mass loading of $\etaMcold = 0.5$. A hot phase mass loading of 0.5 is $\sim 5\times$ larger than expected from SNe ejecta alone. Physically this is consistent with a scenario in which a significant amount of material has been entrained within the subsonic injection region. Relative to the previous, low $\etaM$ examples, the cold clouds in these high $\etaM$ examples are accelerated more rapidly. This enhanced acceleration is in large part due to the significantly larger $\xi$ values and, thus, correspondingly large cloud mass growth rates. The $\xi$ values all begin and remain greater than one out to $\gtrsim 10 $ kpc. The $\xi$ values are larger in these large $\etaM$ examples because the increased wind density, and decreased wind temperature, drive down the cooling times in the clouds' mixing layers, $\tcl$. As a result even the smallest clouds grow, leading to a large decrease in the hot phase mass flux as material is transferred from the hot phase to the cold clouds. The energy of the hot phase falls by more than an order of magnitude as a result of cooling in the mixing layer and cooling in the volume-filling phase itself. The impact of cooling in the volume-filling phase clearly manifests in the sharp decrease in the sound speed of the solutions with low initial cloud masses beyond a few kpc, which corresponds to the cooling radius studied by \citet{Thompson:2016} where $t_{\rm flow} = r/v = \tcoolwind$. These solutions end (marked with an `x' in the $\Mcl$ panel of \autoref{fig:Case_I_high_eta}) if/when the hot phase temperature has dropped to $\Tcl = 10^4$ K, at which point our multiphase description of the winds is no longer valid since the clouds and wind will have fully merged.

The impact of gravity in these solutions is subtle but apparent. The increased mass loading of the hot phase leads to a lower hot phase velocity (\autoref{eq:sonic_radius_values}), which in turn reduces the maximum value of the cold clouds to a few hundred km/s. Because these velocities are comparable to $\vc = 150$ km/s, the hot phase and cold clouds are appreciably decelerated by the gravitational potential. This can be seen in the slight down turn in the cloud velocities after reaching their peak velocity at a few kpc. 

\subsection{The anatomy of multiphase winds; term by term dissections of two solutions}

\begin{figure*}
\includegraphics[width=\textwidth]{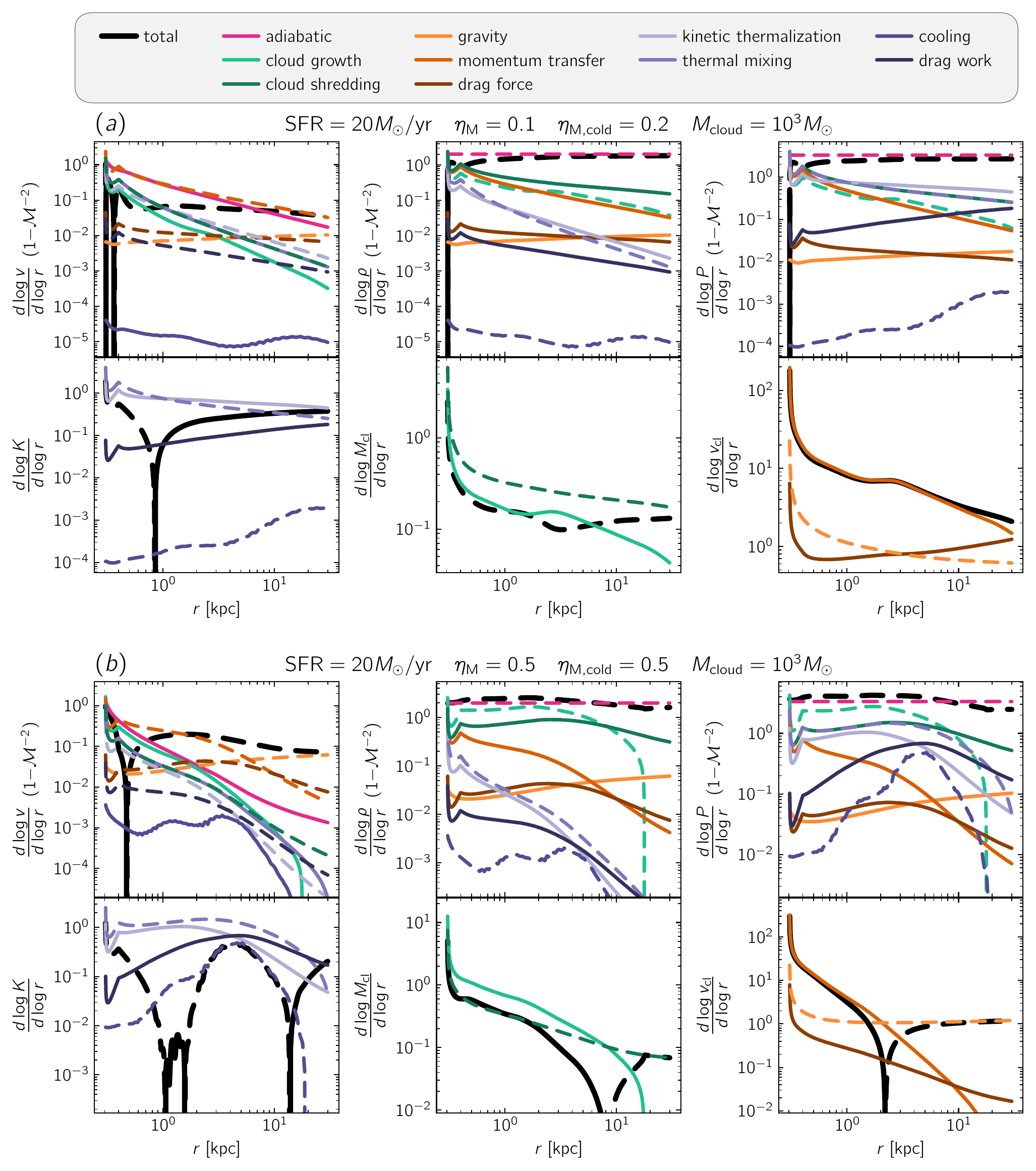}
\caption{Term by term contributions to the logarithmic gradients of $v$, $\rho$, $P$, $K$, $\Mcl$, and $\vcl$ for (a) a wind with SFR = $20 \Msunyr$, $\etaE = 1$, $\etaM = 0.1$, $\etaMcold = 0.2$, and an initial $\Mcl = 10^3\ \Msun$ (red line in \autoref{fig:Case_I_low_eta}) and (b) a wind with SFR = $20 \Msunyr$, $\etaE = 1$, $\etaM = 0.5$, $\etaMcold = 0.5$, and an initial $\Mcl = 10^3\ \Msun$ (red line in \autoref{fig:Case_I_high_eta}). The $v$, $\rho$, and $P$ gradients are multiplied by $(1-\Mach^{-2})$, which is $\sim 1$ beyond a few $\rstar$, for ease of comparison. The colors and labels of each line correspond to Eqs. \ref{eq:velocity_gradient_split_source}, \ref{eq:density_gradient_split_source}, \ref{eq:pressure_gradient_split_source}, \ref{eq:entropy_gradient_split_source},  \ref{eq:Mdotcl_general}, and \ref{eq:vdotcl_general}. Solid and dashed line styles correspond to when the term is positive or negative, respectively. }
\label{fig:Gradients}
\end{figure*}

To better understand how and why the wind properties change as they do we show the term-by-term contribution to the $v$, $\rho$, $P$, $K$, $\Mcl$, and $\vcl$ gradients for two representative examples. The top panels of \autoref{fig:Gradients} show the detailed breakdown of terms for the solution with SFR = 20 $\Msunyr$, $\etaE = 1$, $\etaM=0.1$, $\etaMcold=0.2$, and an initial $\Mcl = 10^3 \ \Msun$, which is also shown with the red lines in \autoref{fig:Case_I_low_eta}. Likewise, the bottom panels of \autoref{fig:Gradients} show the detailed breakdown of terms for the solution with SFR = 20 $\Msunyr$, $\etaE = 1$, $\etaM=0.5$, $\etaMcold=0.5$, and an initial $\Mcl = 10^3 \ \Msun$, which is also shown with the red lines in \autoref{fig:Case_I_high_eta}. 

We focus first on the lower mass loaded example in the top panel of \autoref{fig:Gradients}. The velocity gradient is predominantly set by the balance of adiabatic expansion ($2/\Mach^2$) and momentum transfer between the cold cloud and hot wind ($\propto \rhodotp \vrel$). This momentum transfer, that arises from the addition of slower moving cloud material into the hot fast wind, leads to a net deceleration of the hot phase. The density gradient is predominantly set by the balance of adiabatic expansion ($-2$) and the shredding of clouds that mass loads the hot phase ($\propto \rhodotm$). The impact of cloud shredding/mass loading is largest at small radii, which leads to slight flattening of the density profile, but by $\sim 1$ kpc the density has returned to the familiar $r^{-2}$ solution. The pressure gradient is predominantly set by the balance of adiabatic expansion ($-2 \gamma$) and the thermalization of the cold cloud kinetic energy as the clouds are mixed into the hot phase ($\propto \rhodotp \vrel^2$). As with the density, the deviation from the familiar adiabatic solution is most pronounced at small radii where the transfer rates and relative velocities are highest. The fact that the velocity is set by the momentum source term, the density is set by the density source term, and the pressure is set by the energy source term is intuitive. This straightforward correspondence can, however, be complicated, as we will see below, when the importance of the other terms becomes comparable. 

The $K$, $\Mcl$, and $\vcl$ gradients have fewer terms and are far simpler to understand. The entropy gradient is initially negative as a result of the addition of cold material into the hot wind ($\propto \rhodotp \cscl^2$), but beyond $\sim 1$ kpc the thermalization of the clouds' kinetic energy dominates and the entropy gradient becomes positive. The kinetic energy thermalization heating can be even more significant in solutions with more massive clouds because they are accelerated more slowly and thus maintain larger relative velocities. This is why the large cloud solutions shown in \autoref{fig:Case_I_low_eta} have such high entropy values. Cooling in the hot phase is negligible owing to the relatively low density/low mass loading. Although the cloud mass gradient is negative through the solution for this choice of initial cloud mass the rate at which the clouds lose mass is much slower (reduced by a factor of $>3$ on average) than it would be without the cloud growth term. The clouds, therefore, survive out to much larger radii than would be predicted without the radiative cooling that occurs in their turbulent radiative mixing layers. This cloud growth term, although subdominant in terms of the cloud mass evolution, is responsible for the vast majority of the cloud acceleration. Ram pressure only becomes appreciable at very large distances from the galaxy. In this solution the gravitational deceleration is mostly negligible. 

The fact that the clouds are predominantly accelerated by the transfer of momentum as the clouds grow via mixing layer cooling is reflected in the close correspondence between the increase in cloud velocity and the increase in the cloud metallicity shown in \autoref{fig:Case_I_low_eta}. Comparing the expressions for the rate of change of a clouds velocity and its metallicity in \autoref{eq:vdotcl_general} and \autoref{eq:Zdotcl_general}, respectively, highlights this physical connection between acceleration and metal mixing when other forms of acceleration are subdominant. This is consistent with a wide range of recent studies that have measured the correlation between cold cloud velocity and the degree to which it has been diluted in simulations of galactic winds \citep{Melso:2019,Schneider:2020} and ram pressure stripped galaxies \citep{Tonnesen:2021}.

Although quite idealized, the properties of these first multiphase wind solutions exhibit complex and important behavior. These solutions reproduce key unexplained properties of the complicated and expensive multiphase galactic wind simulations presented by \citet{Schneider:2020}. In particular their simulations and our solutions have flatter density, pressure, and sound speed (or temperature) profiles than a single-phase adiabatic wind, and they have gradually decreasing cold mass fluxes that are matched by gradually increasing hot mass fluxes. The qualitative agreement between their simulations and our model is encouraging and points to the essential role that cold clouds (and their interaction with the hot phase) play in setting the overall properties of galactic winds.  

The bottom panels of \autoref{fig:Gradients} show the term-by-term contributions to the logarithmic gradients of $v$, $\rho$, $P$, $K$, $\Mcl$, and $\vcl$ for a multiphase wind with SFR = $20 \Msunyr$, $\etaE = 1$, $\etaM = 0.5$, $\etaMcold = 0.5$, and an initial $\Mcl = 10^3\ \Msun$. Relative to the $\etaM=0.1$ and $\etaMcold = 0.2$ example shown in \autoref{fig:Gradients} there are several important differences. As in the previous example, in this more mass loaded case, the velocity evolution is dominated by the momentum source terms, but in this case the impact of gravity, which was negligible in the faster, low $\etaM$ case, dominates beyond $\sim 5$ kpc. Likewise, at small radii the cloud velocity $\vcl$ evolution is dominated by the momentum transferred to the clouds from the hot wind due to cooling in the mixing layer, but beyond 2 kpc gravitational deceleration dominates. 

The density of the wind in this example falls off faster than the $r^{-2}$ solution for an adiabatic single-phase wind because the clouds are growing and draining mass from the hot phase well out into the wind. This also leads to a steeper pressure profile. At small radius, the entropy gradient is set by the near balance of thermal energy mixing (of the low entropy cloud gas) and kinetic energy thermalization. At larger distances, however, the wind has adiabatically cooled to the point where radiative cooling in the volume filling phase becomes appreciable. This strong cooling causes the entropy to fall rapidly, but is counteracted by the heating that arises from the drag force.


\subsection{Trends with $\Mcl$, $\etaMcold$, $\etaM$, and SFR} \label{sec:general_trends}

\begin{figure*}
\centering
\includegraphics[width=\textwidth]{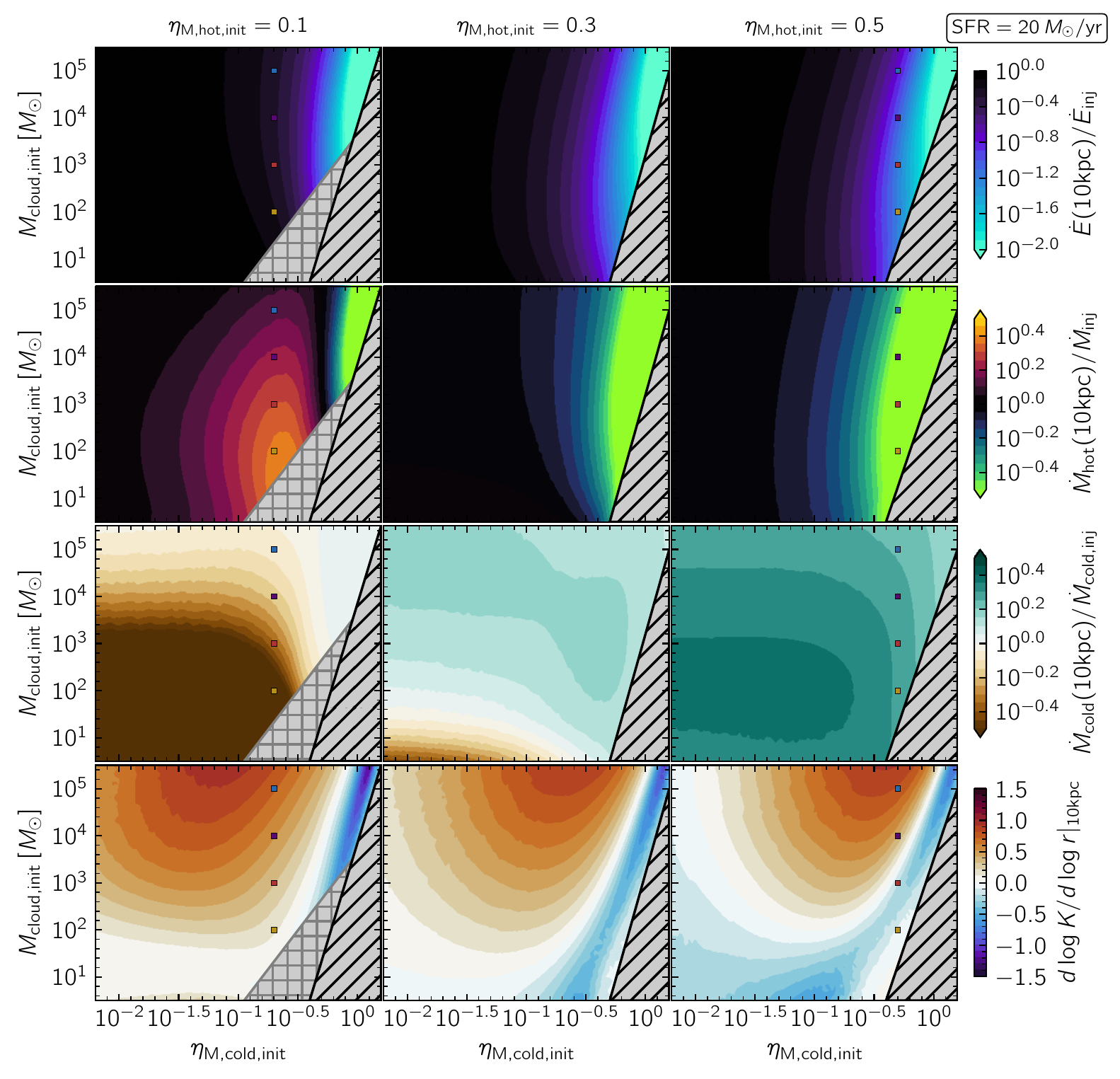} 
\caption{Contours showing the cloud mass $\Mcl$ and cold phase mass loading $\etaMcold$ dependence of the energy flux (top row), hot phase mass flux (second row), cold phase mass flux (third row), and logarithmic entropy gradient (bottom row), all measured at 10 kpc. The fluxes are normalized by the fluxes injected at $\rstar$. From left to right the columns show the results for hot phase mass loading of $\etaM = 0.1, 0.3$, and 0.5. As in the previous examples the SFR = $20 \Msunyr$, $\vc = 150 \kms$, $\Zcl = Z_\odot$, and the hot phase metallicity is $Z = 2 Z_\odot$. The black diagonally hatched region represents solutions in which the wind does not reach to 10 kpc due to the hot phase being entirely drained onto the cold clouds. The gray cross hatched region in the $\etaM = 0.1$ examples mark wind solutions that fail to expand at all due to the rapid mass loading of the hot phase from shredding numerous small clouds, which drives $\Mach < 1$. Increasing $\etaMcold$ leads to an increase in the amount of energy and mass that the hot phase loses and drives the entropy gradient down. For $\etaM = 0.1$ and $\etaMcold \lesssim 0.3$ most choices of $\Mcl$ lead to a decrease in the cold phase mass flux, with more massive clouds surviving for longer. Larger $\etaM$ leads to stronger cooling (in the mixing layers and volume filling phase) and thus more growth of the cold phase. Massive clouds take longer to accelerate, so the heating due to the thermalization of their kinetic energy as they are mixed with the hot phase increases the entropy profile slope. The small colored squares correspond to the examples shown in \autoref{fig:Case_I_low_eta} and \autoref{fig:Case_I_high_eta}.}
\label{fig:Contours}
\end{figure*}

\begin{figure*}
\centering
\includegraphics[width=\textwidth]{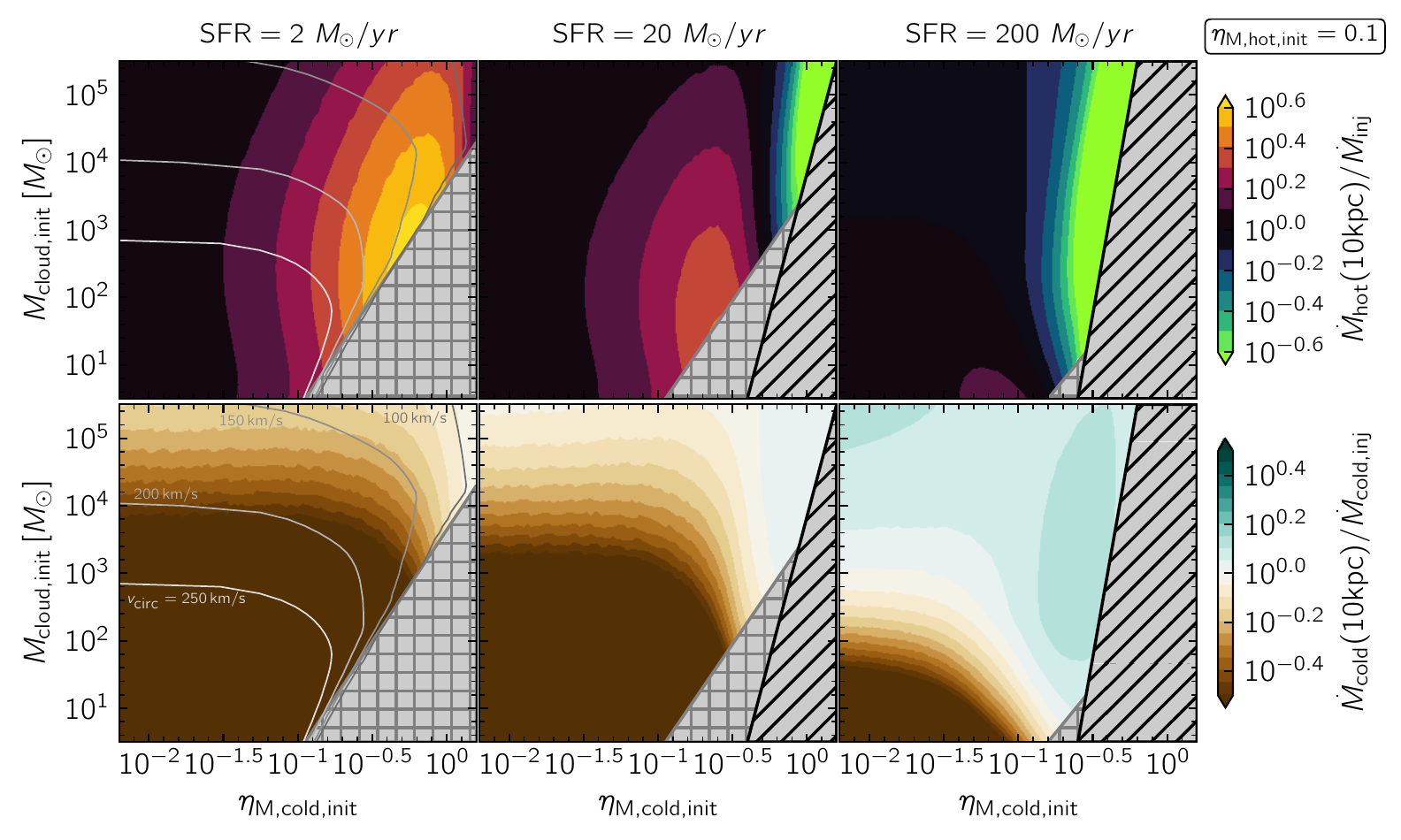} 
\caption{The hot phase mass flux (top) and cold phase mass flux (bottom) at 10 kpc relative to the injected amount as a function of initial cold cloud mass $M_{\rm cl, init}$ and initial cold mass loading factor $\eta_{\rm M, cold, init}$ for $\eta_{\rm M, hot, init} = 0.1$, and SFR = 2, 20, and 200 $\Msunyr$. As in \autoref{fig:Contours}, the diagonally and cross hatched regions show the \emph{poisoned} and \emph{failed} winds, respectively. The higher SFR solutions are more susceptible to radiative cooling (in the TRML) because of the increased density. This leads to more hot phase mass loss as material is transferred to the embedded cold clouds. Lower SFR solutions have lower densities and longer cooling times, which makes the embedded clouds more likely to be shredded, thus mass loading the hot phase at the expense of the cold phase. The thin gray lines on the SFR = 2 $\Msunyr$ panels denote where gravity causes the solution to fail. The effect of gravity is negligible on the higher SFR systems.}
\label{fig:Contours_SFR}
\end{figure*}

The detailed analysis of the solutions presented in the previous section illuminates the physical processes responsible for the wind properties. With this understanding in hand we now turn our attention to the broader trends of the multiphase wind solutions with $\etaMcold$, $\Mcl$, $\etaM$, and SFR. We keep the same simplifying assumptions from the previous examples, in particular that at a given radius all clouds have the same mass and velocity. More realistic and complex assumptions are expected to introduce quantitative not qualitative changes to the measured trends. 

\autoref{fig:Contours} shows several key properties of multiphase wind solutions with SFR = $20 \Msunyr$, cloud mass $\Mcl$ ranging from $10^{0.5}$ to $10^{5.5}\ \Msun$, cold phase mass loading $\etaMcold$ ranging from $10^{-2.2}$ to $10^{0.2}$, and hot phase mass loading of $\etaM = 0.1, 0.3$ or $0.5$. All of the quantities shown in \autoref{fig:Contours} are measured at 10 kpc, the exact choice of which (although somewhat arbitrary) does not change the bulk trends. The first, second, and third rows show the hot phase energy flux, hot phase mass flux, and cold phase mass flux, respectively. Each of these fluxes are normalized by the corresponding injection rate. The bottom row shows the logarithmic derivative of the entropy. In the left and right columns the small colored squares denote the parameters used in the examples shown in \autoref{fig:Case_I_low_eta} and \autoref{fig:Case_I_high_eta}, respectively (omitting the $\Mcl = 10^6 \ \Msun$ point which is beyond the y-range of these plots).

For all choices of $\etaM$ and $\Mcl$, as $\etaMcold$ increases beyond $\sim 0.3$ the hot phase loses an increasingly large fraction of its total energy flux. This energy loss is primarily due to the cloud-wind interaction and the cooling in the TRMLs (and not cooling of the hot phase directly). As a result there is also a corresponding loss of hot mass flux that arises as hot material cools onto the cold clouds. If $\etaMcold$ gets too large the hot component of the wind loses all of its energy to the cold phase, and cools down to the cloud temperature $\Tcl$. This is also reflected in the very negative entropy gradients in this regime. These winds have been effectively \emph{poisoned} by presence of too many cold clouds. The \emph{poisoned} winds are no longer well described by our multiphase wind model and are marked with the black diagonally hatched regions in \autoref{fig:Contours}. In reality, how winds in this regime of parameter space might evolve is unclear with the most likely outcomes being a predominantly cold bursty outflow (i.e. not steady state) or a fountain flow. We will investigate this using simulations in a future work.

When $\etaMcold\lesssim 0.3$ the hot phase suffers much less energy loss and the winds make it out to arbitrarily large radii. There is an appreciable dependence on $\etaM$ in this regime. For large values $\etaM=0.5$ the higher density and lower temperature results in more cooling in the mixing layers (i.e. higher $\xi$) and thus an increase in the cold phase mass flux by a factor of as much as 2.5. Even though $\xi$ increases monotonically with cloud mass the maximal cloud growth occurs for initial cloud masses of $\sim 10^2\ \Msun$ because $\Mdo/\Mcl \propto \Mcl^{-1/3}$. 

For less mass-loaded hot winds with $\etaM=0.1$ the wind density is low and the temperature is high which results in less efficient cloud growth (i.e. lower $\xi$) and a diminution of the cold phase mass flux even for the most massive initial cloud masses $\gtrsim 10^5\ \Msun$. The loss of cold phase mass flux in these low $\etaM$ solutions leads to an increase in the hot phase mass flux. In some cases, however, the rapid addition of a significant amount of mass into the hot phase is sufficient to cause the wind to immediately become subsonic and fail. To a certain extent these \emph{failed} winds are a result of our simplified model and might be avoided by more realistic/elaborate schemes for gradually introducing the cold clouds either within or beyond the subsonic hot wind injection region. However, we keep our simple scheme for now and mark the failed winds with a gray checked region in the left column of \autoref{fig:Contours} because it highlights the real physical effect of \emph{failed} winds due to ``over mass-loading'' the hot phase.

The characteristic acceleration time scale of cold cloud increases for more massive clouds. As a result more massive clouds maintain a substantial relative velocity out to large radii, as can be seen in Figs. \ref{fig:Case_I_low_eta} and \ref{fig:Case_I_high_eta}. The thermalization of this kinetic energy as the clouds are shredded heats the hot phase and is responsible for the increase in the entropy gradient shown in the bottom row of \autoref{fig:Contours} (see the first term in \autoref{eq:entropy_gradient_split_source}). Since the clouds and winds are highly supersonic even a very small amount ($\etaMcold \lesssim 10^{-1.5}$) of large ($\Mcl \gtrsim 10^4 \ \Msun$) cold clouds can increase the entropy gradient substantially. This entropy increase (and the corresponding temperature increase) relative to the standard adiabatic, and single-phase models has clear and important observation implications, particularly at X-ray wavelengths \citep[\eg][]{Lopez:2020}.

\autoref{fig:Contours_SFR} shows the $M_{\rm cl, init}$ and $\eta_{\rm M, cold, init}$ dependence of the hot phase mass flux and cold phase mass flux measured at 10 kpc for multiphase wind solutions with $\eta_{\rm M, hot, init} = 0.1$, and SFR = 2, 20, and 200 $\Msunyr$. The dependence of the hot phase energy flux, and entropy gradient can be inferred from the similar trends shown in \autoref{fig:Contours}. The solutions with higher SFRs have higher hot phase densities which makes them more prone to cooling, both in the hot, volume-filling phase itself and in the mixing layers of the embedded cold clouds. As a result, the hot phases in higher SFR wind solutions begin to lose significant mass and energy by being drained onto the cold clouds at lower values of $\etaMcold$ than in lower SFR wind solutions. Lower SFR multiphase wind solutions, on the other hand, have much lower cooling rates and the embedded cold clouds, even the largest clouds $\gtrsim 10^4 \ \Msun$, are, therefore, much more likely to be shredded and incorporated into the hot flow. 
A result of the clouds being accelerated more slowly in low SFR systems is that the impact of gravity becomes more pronounced. The gray lines on the left column of \autoref{fig:Contours_SFR} show the region above or to the right of which the solutions fail because the cold clouds are not able to overcome gravity. This does not occur in the higher SFR systems. This is consistent with the findings from simulations which have shown that the mass flux able to escape a galaxy decreases for increasing $\vc$, particularly in low SFR systems \citep[\eg][]{Fielding:2018, Kim:2020b}. Overall, the trends with SFR at fixed $\etaM$ (\autoref{fig:Contours_SFR}) are similar to the trends with $\etaM$ at fixed SFR (\autoref{fig:Contours}), which highlights the fundamental dependence on hot phase wind density and mass outflow rate. 

In summary, the structure of a multiphase wind depends primarily on the number of cold clouds ($\etaMcold$), the density of the hot phase ($\etaM$ and/or SFR), and the initial mass of the cold clouds ($\Mcl$). When the number of clouds exceeds a certain threshold ($\etaMcold \gtrsim 0.3$) the hot phase catastrophically cools or decelerates. Below this threshold the magnitude of the (still sizable) impact of the cold clouds on the hot wind increases monotonically with $\etaMcold$. Clouds in winds with a high density hot phase tend to grow due to rapid cooling in their turbulent radiative mixing layers, whereas clouds in winds with a low density hot phase tend to be shredded and mixed into the hot phase. Large clouds ($\gtrsim 10^4 \ \Msun$) survive for longer and the thermalization of their relative kinetic energy provides significant heating to the hot wind.

\section{Discussion} \label{sec:discussion}
Our new model for the steady state structure of a multiphase galactic wind in which the embedded cold clouds can gain or lose mass exhibits a rich diversity of behaviors as the fundamental parameters (SFR, $\etaM$, $\etaE$, $\etaMcold$, and $\Mcl$) are varied. Here we discuss some of the physical implications and uncertainties of our model, the connection of our results to simulations and observations, the missing ingredients and potential extensions of our model, and how to use our framework for a cosmological simulation subgrid model.

\subsection{Physical Implications for Galaxy Evolution}
Building on the recent realization that embedded cold clouds can actually grow and not just lose mass as they are accelerated by a hot wind, our model makes it clear that cold gas in a galactic wind can survive far out into the CGM, and that the presence of this cold gas back-reacts on the hot phase in meaningful ways. These findings require a rethinking of the nature of galactic winds and need to be contextualized in the larger picture of galaxy evolution. 

Galaxies are inefficient at forming stars relative to the total baryon content of their halo as demonstrated by the stellar mass to halo mass relation \citep[\eg][]{Behroozi:2013}. This inefficiency is most pronounced at the low mass and high mass end of the galaxy mass distribution. The inefficiency at the low mass end ($M_{\rm gal} \lesssim 10^{10.5} \Msun$) is generally attributed to effective SF feedback, while at the high mass end ($M_{\rm gal} \gtrsim 10^{10.5} \Msun$) AGN feedback is thought to be primarily responsible. Feedback (either SF or AGN driven) operates in either an ejective and/or preventative mode. Ejective feedback operates by removing material from the star forming ISM of a galaxy, and preventative feedback operates by inhibiting material from making it into the ISM in the first place. Both ejective and preventative feedback are necessary in most galaxy formation theories \citep[\eg][]{Benson:2003}. 

Our model makes clear predictions for the amount of mass that can be ejected from galaxies and carried out by winds powered by SF feedback\footnote{In practice our model could easily be applied to AGN feedback powered winds as well by modifying how the initial conditions are related to feedback processes.}. For fixed SFR, winds with lower $\etaM$ are able to accelerate a larger amount of cold gas (larger $\etaMcold$) without being over-mass loaded and without catastrophically cooling. Likewise, for a fixed hot phase mass loading $\etaM$, the winds emanating from low SFR systems can carry a significant amount of material out of the ISM. The mass flux in these winds is predominantly cold at small radii, but it quickly transitions to being mostly hot as the clouds are shredded by the low density/slowly cooling wind (see \autoref{fig:Contours} and \autoref{fig:Contours_SFR}). The reason for this is that these winds have a lower density, and thus are less susceptible to cooling in the turbulent radiative mixing layers. On the other hand, higher density winds, which can arise when $\etaM$ and/or SFR is large, have lower maximum total mass loading factors because they lose a larger amount of energy to cooling in the mixing layers for even a small number of clouds. The cold phase in these high density winds, however, remains significant and can even continue to grow out to large radii. Although the fraction of mass able to be ejected (i.e., $\etaMtot$) decreases with increasing SFR, the total possible mass flux ($\Mdothot+\Mdotcold$) increases in systems with larger SFRs. It is plausible that the fall off in ejective feedback with increasing SFR may be related to the increasing global galactic efficiency of star formation in dwarf galaxies as galaxy mass increases. All of these outcomes are fundamentally different from single-phase winds.

The mechanisms by which galactic winds can act to prevent future accretion on to a galaxy are less straightforward than ejective feedback and can manifest in a variety of ways. The means by which winds can prevent accretion include, but are not limited to, heating CGM material to prevent it from radiating away its thermal pressure support, and sweeping up CGM material and pushing it away from the galaxy and out into the intergalactic medium (IGM). The ability of a wind to heat the CGM is primarily a question of the wind's energy flux, while sweeping away the CGM depends more on its momentum flux. In both cases, however, it is (likely) the volume-filling, hot component of the wind that is responsible for the preventative feedback rather than the cold clumpy component, since cold clouds subtend a far smaller solid angle than the hot phase, making the cold clouds more likely to pass through the CGM rather than push or heat it (like walking on deep snow in boots rather than snowshoes). Additionally, the cold clouds' interaction with the ambient CGM will be primarily mediated by the TRML transfers, which are less direct than, for example, a volume-filling wind shocking against a hot halo. In light of this, our models also make predictions for the relative importance of preventative feedback from SF driven galactic winds as the galaxy properties change. In particular, the fact that, in winds from low SFR systems, the cold clouds are more susceptible to being shredded and joining the volume-filling phase means that these winds will be effective at preventing accretion in addition to ejecting material. On the other hand, in high SFR systems the clouds are expected to survive longer and even grow, so it is harder for these winds to be as effective at preventing future accretion. This conception is strengthened by our finding that high SFR winds are susceptible to catastrophic cooling, and thus a near total loss of energy flux.

\subsection{What sets $\etaMcold$? A model for the critical value of the cold mass flux.} 
Underpinning all of our models is the choice of SFR, $\etaM$, $\etaE$, $\etaMcold$, and $\Mcl$, which makes it essential to understand what sets each of these parameters and the fundamental uncertainties in our choices. Our fiducial values for these main (and the ancillary) parameters are all motivated by simulations and observations to a certain degree. For example, we used the \texttt{twind} package that conveniently reports how $\etaM$, $\etaE$, and $\etaMcold$ depend on SFR in the TIGRESS simulations \citep{Kim:2020b}, as the motivation for choosing $\etaE = 1$, and $\etaM = 0.1$, and for choosing the range of $\etaMcold$ values we explored. Likewise, we focused on SFR values that are in the range of what was simulated by \citet{Schneider:2020}. However, there is a benefit to a somewhat more first-principled approach. It is easy to convince one's self that SFR, $\etaM$, and $\etaE$ are closely related and set by the intricacies of star formation and SNe, which are beyond the scope of this model. The question then becomes what sets $\etaMcold$ and $\Mcl$. 

The model we have presented here, in which there is a single initial cloud mass and that all clouds evolve identically is clearly an oversimplification. In reality there should be a distribution of cloud masses and velocities. Nevertheless it is worthwhile to consider what sets the cloud mass distribution and if there are characteristic values. This is fundamentally a question of ISM physics and how turbulence and other processes determine the giant molecular cloud (GMC) mass function. Observations and theory predict characteristic upper cut off GMC masses of $\sim 10^6 \ \Msun$ with smaller clouds being more common \citep[\eg][]{Williams:1997,Hopkins:2012c}, which is consistent with the range we studied. We are currently extending our formalism to include a full cloud distribution (Anthony Chow et al., in prep). 

\begin{figure*}
\centering
\includegraphics[width=\textwidth]{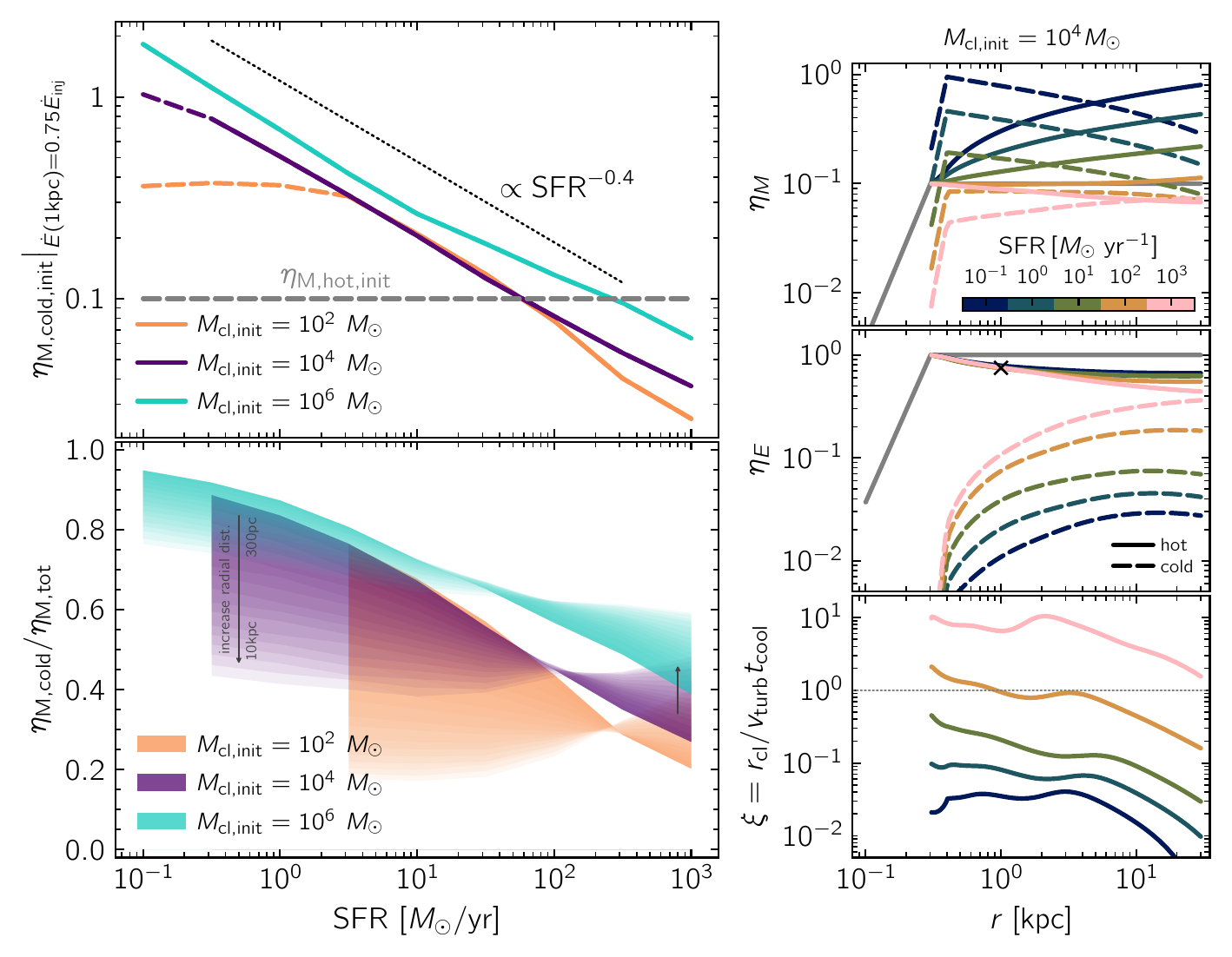} 
\caption{(\textbf{Upper Left}) The SFR dependence of the value of the initial $\etaMcold$ such that the hot phase has retained 75 \% of its energy flux by the time it reaches 1 kpc for winds that start with $\etaM=0.1$. The different colored lines correspond to different initial cloud masses. The dashed portion of the colored lines show where no such solution is possible because the winds become over-mass-loaded and \emph{fail} at a value of $\etaMcold$ less than what would be necessary. The critical value of $\etaMcold$ scales proportional to SFR$^{-0.4}$. The SFR scaling is insensitive to the arbitrary choice of the energy loss percentage, with the only change being a marginal decrease in $\etaMcold$ when there is a lower target energy flux reduction. (\textbf{Lower Left}) The SFR dependence of the fraction of the total mass flux carried cold phase. The gradient in the shaded region shows the radial dependence of the cold mass flux fraction with small radii shown with opaque colors giving way to larger radii shown with increasing transparency. The \emph{failed} wind SFR range is omitted. For all choices of initial $\Mcl$ the cold fraction decreases with radius when SFR $\lesssim 100 \ \Msunyr$, and increases with radius for larger SFRs. (\textbf{Right}) The radial profiles of the mass loading $\eta_{\rm M}$, the energy loading $\etaE$, and the ratio of the mixing time to the cooling time in the turbulent radiative mixing layer $\xi$ for several of the winds with an initial $\Mcl = 10^4\ \Msun$ and $\etaMcold$ consistent with the left panels. The winds from higher SFR systems have higher density and thus large $\xi$ values, which is why they cannot sustain as large of a cold mass flux without losing more energy. The X in the central right panel marks are assumed energy loss fraction at 1 kpc.}  \label{fig:eta_M_edot_SFR}
\end{figure*}

How the value of the cold cloud mass flux, parameterized by $\etaMcold$, is determined is unclear. In simulations that resolve these processes, and presumably in reality, cold clouds in the ISM are swept up by the hot superbubbles that expand violently in the wake of a series of SN explosions. As these bubbles reach the scale height of the gas disk, they blow a chimney through the galaxy and launch the shell of material they have swept up, as well as the cold clouds that have survived, incorporating material ablated from clouds, and the SN ejecta out into the wind \citep[\eg][]{MacLowMcCray:88,Koo:1992, Fielding:2018}. After breakout, there is a continued flux of cold clouds into the chimney which may add to the available cold cloud mass flux. A key behavior of this process is that not all of the cold material the hot wind interacts with is necessarily swept up and launched out in the wind. This is because the ISM is highly porous and so when the hot superbubble runs into a large cold cloud that it is not able to immediately overwhelm it will instead go around it through a lower density channel. 
This may lead to a self-regulated $\etaMcold$ value since, out of the effectively inexhaustible supply of cold ISM material, a superbubble that manages to break out will sweep up only as many cold clouds as it can without radiating away all of its energy.
Although this is, of course, an oversimplification of the complex processes regulating superbubble breakout and cloud entrainment, it is in line with the picture emerging from both ISM patch simulations and cosmological zoom-in simulations. These simulations show that superbubbles in dwarf galaxy conditions tend to sweep up large swaths of the ISM and breakout nearly spherically, whereas in larger/more massive systems the breakout happens in narrow chimneys that feed biconical outflows and carry only a small fraction of the cold material out of the galaxy \citep[][]{Muratov:2015,Tanner:2016,Gutcke:2017,Hafen:2019,Nelson:2019,Peroux:2020,Mitchell:2020,Pandya:2021}. 

These processes are not directly captured by our model which requires the cold phase mass flux to be specified, however, by assuming that $\etaMcold$ is set by a self-regulating process that ensures the wind sweeps up as much material as it can without losing too much energy we can get a sense for how such a process might work. \autoref{fig:eta_M_edot_SFR} shows the result of such an experiment in which we assume that the value of $\etaMcold$ is as large as possible without causing the wind to lose more than 25\% of its initial mass flux by the time it has reached 1 kpc.\footnote{The qualitative behavior of this experiment is largely insensitive to the exact choices for the location and magnitude of the target energy flux diminution.} The premise behind this experiment is further supported by the finding that, in highly realistic ISM patch simulations spanning a wide range of $\SigmaSFR$, values of the energy loading factor of the hot phase $\etaE$ are of order unity and essentially independent of $\SigmaSFR$ \citep{Li:2020a,Kim:2020a}.  \autoref{fig:eta_M_edot_SFR} demonstrates that multiphase galactic winds powered by lower SFRs are able to sustain much higher total mass loading factors than winds powered by higher SFRs. The upper left panel of \autoref{fig:eta_M_edot_SFR} shows the SFR scaling of the value of the critical initial $\etaMcold$ such that the hot phase has retained 75 percent of its energy flux by 1 kpc in the case when $\etaM = 0.1$. Winds from low SFR systems can have $\Mdotcold \gtrsim 10 \Mdothot$ at the base of the wind where they are launched. The cold phase mass loading factor scales roughly proportional to SFR$^{-0.4}$, which is quite similar to the $\etaMcold \propto \Sigma_{\rm SFR}^{-0.44}$ scaling found in the TIGRESS ISM patch simulations \citep{Kim:2020a}. 

Although the cold mass flux can dominate at the base of a wind powered by a relatively low SFR, by several kpc the cold clouds will have been mostly shredded and added to the hot flow. This is shown in the lower left panel of \autoref{fig:eta_M_edot_SFR}, which shows how the fraction of the mass flux carried by the cold phase changes from small radii (opaque colors) to large radii (transparent colors). Thus, our model predicts that low SFR winds can be highly mass loaded and will tend towards being predominantly hot and single-phase as they move out from the galaxy. Winds from high SFR systems will, on the other hand, have lower total mass loading factors, but they will maintain a multiphase nature with an appreciable cold phase out to larger radii. This is consistent with  \autoref{fig:Contours_SFR}. The right panels of \autoref{fig:eta_M_edot_SFR} show the radial profiles of the hot and cold mass and energy loading factors (top and middle) and $\xi$ (bottom) for winds with the critical $\etaMcold$, an initial $\Mcl = 10^4 \ \Msun$, and a range of SFRs. The large $\xi$ values in the high SFR winds is a reflection of the fact that high wind densities makes the winds susceptible to significant energy loss per cloud interaction, and thus they can only sustain sweeping up a much smaller amount of cold material than the low SFR winds that have low densities and correspondingly low $\xi$ values. The radial profiles further demonstrate that although there are fewer cold clouds in high SFR winds, they survive to large radii, while the much more numerous cold cloud population in low SFR winds are rapidly destroyed. 

The prediction from this simple experiment, that there are higher mass loading and larger cold fractions at small radii in lower SFR systems, is consistent with the range of high resolution simulations including global galaxy simulations \citep{Fielding:2017b, Schneider:2020}, ISM patch simulations \citep{Fielding:2018, Kim:2020a}, and cosmological zoom-in simulations \citep{Pandya:2021}. Moreover, this is in line with observed inverse scaling of mass loading and SFR \citep[\eg][]{Heckman:2015}. 

\subsection{Simulations}
We now discuss our findings in the context of existing simulations of multiphase galactic winds. We focus on three categories of simulations that specifically attempt to resolve the multiphase structure of winds, which are ({i}) isolated galaxy simulations, ({ii}) ISM patch simulations, and ({iii}) cosmological simulations. 

There are relatively few isolated galaxy simulations that have been analyzed in the context of their radial profiles as they expand out in the surrounding medium, but those that have been published provide an excellent point of comparison for our model. The preeminent example is the CGOL simulation \citep{Schneider:2020}, which modeled an M82-like galaxy with uniformly very high (5 pc) resolution over a large ($10 \times 10 \times 20$ kpc$^{3}$) volume. The wind is powered by the feedback from massive star cluster formation, which was set to correspond to a SFR $=20 \ M_\odot$/yr at early times and SFR $=5 \ M_\odot$/yr at later times. The feedback energy was injected into the ISM and the properties of the resulting wind, such as $\etaE$, $\etaM$, and $\etaMcold$, develop self-consistently. Note, however, that the ISM in this simulation is single-phase due to the imposed temperature floor at $10^4$~K, this is noticeably different from the state-of-the-art ISM patch simulations which allow cooling down to $\sim 10$~K and generically have a highly multiphase ISM structure. This is likely responsible for key differences between the two approaches, particularly in the properties of the cold phase of the winds. Beyond the ISM the assumption of a $10^4$~K floor is likely a good assumption because the lower densities and thus more prominent impact of photoionization. The \cite{Fielding:2017b} global isolated galaxy simulations, which are also well-suited for comparison albeit with radially decreasing resolution, of smaller galaxies, and did not split the analysis by temperature, also used a $10^4$~K temperature floor. 

There are several salient properties of the \cite{Schneider:2020} and \cite{Fielding:2017b} isolated galaxy simulations that our model both reproduces and sheds light on the underlying physical processes. In all cases these simulations radial profiles exhibit appreciable changes relative to the expectations from single-phase adiabatic and/or radiative cooling solutions (see \autoref{fig:CC85_Cooling_and_Gravity}). Specifically the temperature, density, and pressure of the hot phase fall off more slowly with radius than expected, and the radial velocity is lower than expected. This is a common occurrence in our models (see \autoref{fig:Case_I_low_eta} and \autoref{fig:Case_I_high_eta}), and is a result of the exchange of mass, momentum, and energy between the cold and hot phases (see \autoref{fig:Gradients}). In effect, the presence of the cold clouds within the hot wind give rise to a radially distributed set of source terms that leads to these changes to the radial profiles. 

Furthermore, \cite{Schneider:2020} demonstrate that during the SFR = 20 $M_\odot$/yr portion of their simulation the cold phase mass flux is comparable to the hot phase mass flux at small radii. The cold phase mass flux, however, decreases with radius, while at the same time the hot phase mass flux increases (see their Fig. 8). During this part of their simulation, the hot phase has $\etaE \sim 1$ and $\etaM \sim 0.1$ at the base of the wind. Determining the cold phase mass flux at the base of the wind is complicated somewhat by the biconical nature of the wind and the distributed injection of the cold clouds, but mass conservation indicates an initial value of $\etaMcold \approx 0.15-0.2$ (this is consistent with the mass flux profile averaged over a wider opening angle, E. Schneider private communication). These values of the SFR, $\etaE$, $\etaM$, and $\etaMcold$, closely match the conditions of our multiphase wind solutions shown in \autoref{fig:Case_I_low_eta}. The metallicity and $\vc$ are also in close agreement between our solution and their simulation. Despite the numerous simplifications inherent to our model the detailed behavior of the mass (and energy) flux profiles agrees with the simulation results remarkably well. For any initial cloud mass less than $\sim 10^6\,\Msun$ the cold phase mass flux decreases with radius, which further mass loads the hot phase. The closest agreement between the simulated results and our model is with an initial cloud mass of $10^3 \ \Msun$, which is consistent with the simulation cloud sizes of a few 10s to 100s pc. The agreement between the simulations and our model, in terms of the mass fluxes, as well as the radial profiles, is an encouraging indication that our model captures the essential behavior of multiphase galactic winds and can be used to understand the dynamics of these complex and expensive simulations (and observations!). Future detailed comparisons of these, or similar, simulations could open a valuable window into the inner workings of galactic winds.

Simulations of roughly 1 kpc patches of the ISM have been used for decades to study star formation, winds, and the link between the two. There is immense diversity in the simulations and the conclusions that have been drawn from them. There is, however, a coherent picture emerging from these simulations that the hot component of SF driven winds generally have a low mass loading $\etaM \sim 0.1$ and a high energy loading $\etaE \sim 1$ \citep[see][and references therein]{Li:2020a}, which motivated our adoption of these as our fiducial values. Furthermore, there is a (rough) consensus settling on the idea that the cold phase carries significantly less energy, and has a mass loading that can be $\etaMcold \gg 1$ in low surface density systems (either SFR surface density $\SigmaSFR \lesssim 10^{-2} M_\odot  {\rm kpc}^{-2} {\rm yr}^{-1}$, or gas surface density $\Sigmagas \lesssim 10 M_\odot  {\rm pc}^{-2} $), and that $\etaMcold$ decreases for increasing surface density \citep[\eg][]{Kim:2020b, Pandya:2021}, which is consistent with the results of our model. 

Cosmological simulations also provide a useful point of comparison for our model. Most large volume cosmological simulations have resolution that is too low to allow for a meaningful comparison since it would be impossible for them to resolve even the largest cold clouds. Some cosmological zoom-in and high resolution cosmological simulations, however, have sufficiently small particle masses that they can marginally resolve the larger cold clouds we considered. Recently, \cite{Pandya:2021} characterized the multiphase nature of the outflows in the FIRE-2 cosmological zoom-in simulations, which have resolution of $\sim 7\times10^3\, \Msun$. Their analysis focused on how the mass, momentum, and energy flux is carried by material in different temperature bins, and how this changes with host galaxy properties. In broad agreement with ISM patch and isolated galaxy simulations they found that the outflow loading factors are smaller in more massive systems and that they scaled roughly as $\SigmaSFR^{-0.5}$. The fraction of outflows carried by hot gas ($T>10^5$~K) increases with galaxy mass, and the fraction carried by colder gas ($T<10^5$~K) decreases with galaxy mass. In the TNG50 simulation galactic winds from lower mass galaxies (where SF feedback is dominant over AGN feedback) are multiphase and exhibit many of these same properties, despite having $\gtrsim 10\times$ larger mass resolution ($8.5\times10^4\ \Msun$) and a very different feedback model \citep{Nelson:2019}. The agreement between these disparate cosmological simulations points to the robustness of the galactic wind properties. Furthermore, these cosmological simulation winds fit neatly into the qualitative picture that our model paints.

\subsection{Observations}

Comparisons between our model and observations must be limited to general (as opposed to detailed) properties given the highly idealized nature of our model. Nevertheless, these rough comparisons can provide valuable insight into the validity of our model and the processes that regulate the structure of observed systems. We briefly discuss observed mass flux and velocity trends in the context of our findings, as well as model predictions of surface brightness, velocity gradient, and cloud property profiles in the context of observed systems particularly the iconic local galactic wind systems M82 and NGC253. We also discuss potential implications for CGM observations.

\begin{figure*}
\centering
\includegraphics[width=\textwidth]{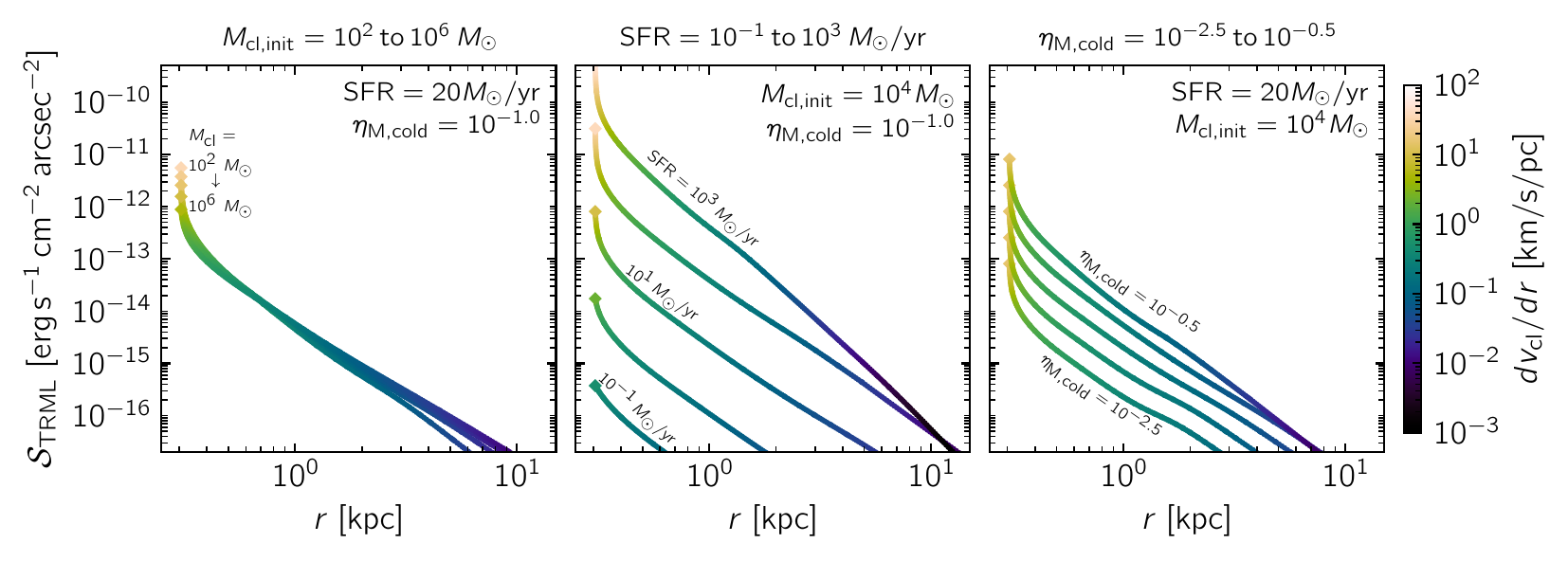} 
\caption{Surface brightness as a function of radius for winds with $\etaE = 1$, $\etaM=0.1$, $\Mcl$ ranging from $10^2$ to $10^6$ (left), SFR ranging from $10^{-1}$ to $10^3$ $\Msun /$yr (middle), and $\etaMcold$ ranging from $10^{-2.5}$ to $10^{-0.5}$ (right). The line color denotes the cold cloud velocity gradient $d \vcl /dr$. The diamond symbol denotes the beginning of the lines. The surface brightness profiles and velocity gradients are nearly independent of cloud mass, whereas the normalization of the flux profiles increases with SFR and $\etaMcold$. \label{fig:FluxProfile}}
\end{figure*}

Down-the-barrel absorption line observations provide one of the most readily accessible observational probes of the properties of galactic winds over a wide range of cosmic time. This method studies the absorption of the host galaxy itself by material in the wind, generally neutral or photoionized metals or H\textsc{i}. This method allows for the robust measurement of the maximum  and centroid outflow velocity and the column density of ion (or ions) that are observed. Rough estimates for the mass, momentum, and energy fluxes are often made by modeling the ionization fraction to convert from ion to total column density, and by assuming some characteristic radius of the outflow. Some notable results from such observational exercises relevant to our model are that the wind velocity increases as $v \propto$~SFR$^{\sim0.2-0.3}$ \citep[\eg][]{Martin:2005,Chisholm:2015}, and that the cold (using our definition of cold to mean $\sim 10^4$K material) phase mass loading $\etaMcold$ decreases with increasing SFR \citep[\eg][]{Rupke:2005b,Heckman:2015}. Both of these observational findings are consistent with the multiphase wind solutions produced by our model. The increase in velocity with increasing SFR is explained in our model by the fact that higher SFRs lead to more efficient cooling in the mixing layers of the cold clouds, and thus more rapid acceleration of the cold clouds (see \autoref{fig:Case_I_high_eta}, \autoref{fig:Contours_SFR}, and \autoref{fig:FluxProfile}). The observed decrease of $\etaMcold$ with increasing SFR is, likewise, explained in our model by the fact that the efficient growth and acceleration of cold clouds in high SFR systems limits the total amount of cold clouds the wind can sustain without completely cooling (see \autoref{fig:Case_I_high_eta}, \autoref{fig:Contours_SFR}, and \autoref{fig:eta_M_edot_SFR}). 

The extensive multi-wavelength observation of the best local galactic wind system M82 gives us access to more detailed and radially resolved properties of the wind. \cite{Lopez:2020} modeled the X-ray emission as a function of distance from the central starburst of M82 and revealed that the temperature of the hot phase falls off with radius much slower than expected from adiabatic or radiative single-phase galactic wind models \citep[\eg][]{ChevalierClegg:1985, Thompson:2016}. This slow decline of the hot phase temperature with radius is similar to what was seen in the CGOLs simulations \citep{Schneider:2020}. This is one of the most compelling existing observational findings for our model since in many cases, particularly systems with properties similar to M82, our model predicts a significant flattening of the temperature profile, as shown in \autoref{fig:Case_I_low_eta}. Our model predicts that in systems like M82, which have relatively high SFRs and low $\etaM$, the embedded cold clouds will be shredded. A direct result of the shredding of the cold clouds is the flattening of the temperature profile, which arises from (i) heating due to the thermalization of the cold cloud kinetic energy as they are shredded, and (ii) reduced expansion as the hot phase is mass loaded and decelerated. Our prediction that M82's cold clouds will be shredded is consistent with a recent observation of CO emission from cold clouds out to several kpc \citep{Krieger:2021}. These observations indicate that the median cloud mass falls from $\sim10^{5.5} \ \Msun$ at $r \approx 0.5$ kpc to $\sim10^{4.5} \ \Msun$ at $r \approx 2$ kpc. This clear decrease in cloud mass is consistent with our model predictions.

Multi-wavelength emission observations of local systems provides an information rich probe into the detailed nature of galactic winds. Although in its current form our model cannot be used to forward model such observations because it does not track the full phase structure and ionization state of the wind\footnote{In a future work we plan to extend our framework to include a full model for the phase structure and ionization state.}, we can make approximate predictions for the total emission as a function of radius. \autoref{fig:FluxProfile} show predictions for the surface brightness as a function of radius for winds with $\etaE = 1$, $\etaM=0.1$, $\Mcl$ ranging from $10^2$ to $10^6$, SFR ranging from $10^{-1}$ to $10^3$ $\Msun /$yr, and $\etaMcold$ ranging from $10^{-2.5}$ to $10^{-0.5}$. The line color denotes the cold cloud velocity gradient $d \vcl /dr$. The surface brightness is approximated by $\mathcal{S}_{\rm TRML} = r \ncl \edotcool / 4 \pi$, where $\ncl \edotcool$ is the volumetric cooling rate in the turbulent radiative mixing layers of all the clouds at that radius (see \autoref{eq:epdotcool_general}). This emission will be spread out from X-rays to the infrared, exactly how it is emitted requires knowledge of the phase structure of the TRML. Current theoretical estimates indicate that the majority of the cooling peaks around $2\times 10^4$~K and the shape of the phase distribution remains similar across a range of conditions \citep[\eg][]{Kanjilal:2021, Abruzzo:2021, Tan:2021b}. We can therefore assume that the majority of the emission is in Lyman $\alpha$ and that roughly 10 percent is in H$\alpha$ \citep{Osterbrock:1989}. Despite the clearly very rough nature of this approximation it agrees fairly well with existing observations and makes testable predictions. Existing observations of the H$\alpha$ flux from the inner wind of M82 and NGC253 are on the order of $10^{-15}$ erg s$^{-1}$ cm$^{-2}$ arcsec$^{-2}$ \citep{McKeith:1995,Westmoquette:2007,Westmoquette:2009,Westmoquette:2011}, which is similar to what our model predicts. Furthermore, \autoref{fig:FluxProfile} makes a clear prediction of a correlation of the cold cloud velocity gradient (which encodes the cloud acceleration) and the emission. This is testable with emission estimates coming from, for example, H$\alpha$, [N\textsc{ii}], and [O\textsc{iii}], which all probe material in the turbulent radiative mixing layer, and with velocity gradient measurements coming from, for example, CO measurements, which probe material in the cold clouds. The ability to make these measurements in more systems is becoming increasingly possible with surveys using instruments such as KCWI \citep[\eg][]{Rupke:2019} and MUSE, and using the radio telescope ALMA. Observational verification of this prediction would provide a robust validation of the fundamental concept of TRML entrainment that is the basis of our model. 

Quasar absorption line observations at large impact parameters ($\lesssim 150$ kpc) have conclusively shown that low-ionization metal species are abundant in the CGM at low and high redshift, and around both massive and dwarf galaxies \citep[\eg][]{HWChen:2010, Prochaska:2011, Werk:2014, Johnson:2017, Rudie:2019}. The presence of so much low-ionization gas is consistent with there being as much as $\sim 10^{10} \ \Msun$ of cold ($T\sim 10^4$ K) gas in the CGM of low-redshift Milky Way-like galaxies \citep{Prochaska:2017}. The source of this large reservoir of cold CGM gas is debated, but galactic winds are a likely contributor. We have demonstrated that in many conditions cold clouds can survive and even grow in galactic winds. Winds may, therefore, be responsible for an appreciable fraction of the observed cold CGM. This is supported by the finding that low-ionization CGM absorption is enhanced in starburst galaxies that have strong winds \citep{Heckman:2017}.

\subsection{Cosmological Simulation Subgrid Model}

Much of the framework we have introduced here can be used as a simulation subgrid model for the difficult to resolve cold gas. This would be particularly useful for isolated galaxy and cosmological simulations in which it is computationally infeasible to simultaneously resolve both the large scale structure of the CGM and the small scale structure of the cold clouds and their turbulent radiative mixing layers. In an upcoming work, as part of the SMAUG\footnote{\href{https://www.simonsfoundation.org/flatiron/center-for-computational-astrophysics/galaxy-formation/smaug/}{Simulating Multiscale Astrophysics to Understand Galaxies}} collaboration, we will introduce a subgrid model that uses the framework presented here in the code AREPO \citep{Weinberger:2020}. Roughly speaking this subgrid model works by probabilistically launching hot and cold wind particles. The hot wind particles have high specific energy and are decoupled from the hydrodynamics after being launched until they have left the star forming ISM, at which point they deposit their mass, momentum, and energy all at once. This is similar to the hydro decoupled scheme currently used in the TNG simulations \citep{SpringelHernquist:2003}. The cold wind particles, on the other hand, will be longer lived and will exchange mass, momentum, and energy with their host cells according to the model we have presented here. 

Recently, a similar subgrid model, called the PhEW model, was developed and used in cosmological simulations \citep{Huang:2020,Huang:2021}. As in our model, the PhEW model represents the cold clouds with a subgrid model that allows them to deposit mass into their host cells over the life of the cloud.  The major differences between our model and the PhEW model are that (i) the PhEW model only launches cold, low specific energy wind particles, (ii) in the PhEW subgrid model the cold clouds can only lose mass, and can never gain mass from the hot phase, and (iii) the PhEW cold clouds have a much more detailed model for their evolution. The PhEW subgrid model is founded upon a thorough and complex model for the detailed evolution of a cold cloud moving relative to a hot wind that is based primarily on the \cite{Bruggen:2016} conductive cloud crushing simulations. These simulations include radiative cooling but do not exhibit the cloud growth that is present in other simulations because of the range of their parameter study and/or the size of their computational domain.  Their cold cloud evolution includes many effects we omitted in \autoref{sec:cloudevo}, such as, a Mach number dependence, the shock structure, a time dependent elongation, and conductive mass loss, which allowed them to capture behaviors we cannot. On the other hand, our choice to adopt a simpler cloud evolution model, with the notable addition of a cloud growth channel, allowed us to focus on the back reaction and interplay of the cold clouds and the hot wind. In the future, much will be gained by combining aspects of our complementary approaches. 

\subsection{Missing Ingredients and Model Extensions}

There are numerous physical processes and relevant scenarios that we have not included in our model. Here we briefly discuss a non-exhaustive subset of these missing ingredients as well as some model extensions.

\emph{Magnetic fields}: We have not taken into account the impact of magnetic fields on the hot wind, the cold clouds, or the interaction between the two phases. The presence of magnetic fields, even with $\beta = P_{\rm th} / P_{\rm mag}$ significantly larger than 1, can modify the transfer of material between the cold and hot phase. Magnetic fields have been shown to slow the destruction of cold clouds being ablated by a hot wind and enhance the drag \citep{Dursi:2008,McCourt:2015}. In the presence of strong cooling, magnetic fields can lead to an appreciable reduction in cloud mass growth at early times ($\lesssim {\rm few\, } \tcc$) \citep{Gronnow:2018}, but by late times the mass growth is comparable to or greater than the pure hydro case \citep{Gronke:2020}. The cooling and condensation process leads to a magnetic field build up in the cold clouds which can eventually lead to a significant thermal pressure imbalance between the phases, while maintaining a total pressure balance \citep{Nelson:2020}. Futhermore, magnetic fields carried from the ISM by galactic winds have been recently shown to provide a significant source of magnetization of the CGM \citep{vandeVoort:2021}. All together it is unlikely that magnetic fields will change the qualitative results of our model, but the presence of magnetic fields in multiphase galactic winds are likely to result in many interesting physical and potentially observable changes.

\begin{figure}
\centering
\includegraphics[width=0.5\textwidth]{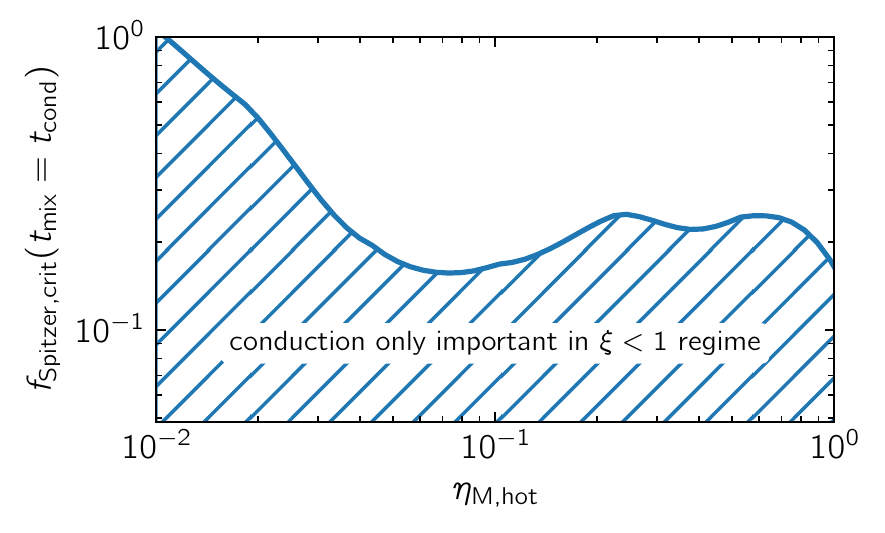} 
\caption{ The blue line shows the critical Spitzer conduction suppression factor value below which a cold cloud would be in the mass loss regime prior to the onset of appreciable conductive evaporation. This is independent of cloud size, and varies by only a few percent when SFR is varied by 5 orders of magnitude. The suppression of conduction due to resonant scattering of electrons off of whistler waves leads to a supression of at least 0.1. It is, therefore, unlikely that conduction will ever dominate in the $\xi > 1$ regime. \label{fig:Supression_Factor}}
\end{figure}

\emph{Conduction}: The inclusion of thermal conduction is a potentially major omission from our model since it can significantly change the evolution of a cold cloud in a hot wind. Conductive evaporation can lead to a net mass loss from small embedded cold clouds \citep{Cowie:1977}. When the cold cloud also has a velocity relative to the hot phase conductive evaporation can, in some cases, dominate over turbulent shredding. Despite this, conductive cloud crushing experiments have demonstrated that sufficiently large clouds can grow under the combined influence of conduction, turbulent mixing, and cooling \citep{Armillotta:2016}. Recent theoretical models, numerical simulations, and experimental data have demonstrated that the canonical \citet{Spitzer:1956} model overestimates the strength and temperature dependence of the conductivity of ionized plasmas. This recent work has found that electrons are resonantly scattered by self-generated whistler waves, which suppresses the heat flux by at least an order of magnitude relative to the Spitzer estimate \citep[\eg][]{Roberg-Clark:2016, Komarov:2018, Drake:2020, Meinecke:2021}. The impact of conduction on the results of our model would be most pronounced if it were strong enough to make clouds that would otherwise be in the growing regime transition to the mass loss regime. Conduction causes clouds to evaporate when the Field length ($\lambda_{\rm F} = \sqrt{\kappa T / n^2 \Lambda}$) \citep{BegelmanMcKee:1990} is larger than the cloud size. We can therefore define a critical Spitzer conduction suppression factor $f_{\rm Sp, crit}$ such that $\lambda_{\rm F}/\rcl > 1 $ only when $\xi < 1$. If conduction were suppressed by at least this much then the inclusion of conduction would only slightly modify the mass loss rate of the already shredding clouds. To calculate this critical $f_{\rm Sp, crit}$ we make a conservative estimate of $\lambda_{\rm F}$ by using the hot phase temperature and the mixing layer cooling time using the properties of the wind just beyond the sonic radius, where conduction is most likely to be important ($f_{\rm Sp, crit} = \vturb^2 \tcl^2 n^2 \Lambda / \kappa T$). The result of this calculation, which are shown in \autoref{fig:Supression_Factor}, is that the true conductivity need only be suppressed by $\lesssim 0.2$ relative to the Spitzer estimate. This value is nearly independent of SFR, and has only a very weak depedence on the hot phas mass loading $\etaM$. It is therefore likely that the inclusion of conduction would not qualitatively change our results, especially if whistler-wave suppression were included. In a future work we plan to adopt some of the detailed conductive evaporation modelling developed by \cite{Huang:2020}. Lastly, although conduction is unlikely to change the overall dynamics of a multiphase wind it is crucial for setting the phase structure of the intermediate temperature gas and thus for making detailed predictions for emission and absorption.

\emph{Self-shielding, low temperature cooling, self-gravity, and non-equilibrium ionization}: Our model currently assumes that the cold clouds are all at $10^4$ K, however at a certain point large clouds should reach high enough column densities to shield themselves from external photoionization sources. In the self-shielding regime clouds will be able to cool to significantly lower temperatures ($\sim 10^2$ K) and higher densities, and in some cases the clouds own self-gravity will become significant \citep{Girichidis:2021}. This will have important observational implications as well as lead to modifications to the cloud-wind interaction \citep{Farber:2021}. In principle it will be straightforward to modify our current cloud-wind interaction to include the ability for high column density clouds to cool to lower temperatures. Likewise, we have throughout assumed that the ionization state of the volume-filling wind, and the mixing layer is in equilibrium, which will not hold generically and can lead to appreciable changes to the dynamics and observable properties of both phaes \citep[\eg][]{Ji:2019,Gray:2019a}

\emph{Cosmic rays}: It is well established that cosmic rays have the potential to play an important role in driving galactic winds and heating the CGM \citep[\eg][]{Socrates:2008, Salem:2014, Jana:2020, Butsky:2018, Quataert:2021a,Quataert:2021b}. There is, however, significant uncertainty in details of cosmic ray transport, which can lead to dramatically different predictions for their impact on galaxy formation \citep[\eg][]{Hopkins:2021a}. Recent cosmological simulations including cosmic rays indicate that cosmic ray driven winds become important at low redshift ($z<1$) in somewhat massive halos ($M_{\rm halo} > 10^{11} \Msun$) \citep{Hopkins:2021}. Thus, our model, which neglects cosmic rays, is, at a minimum, relevant during the peak of cosmic star formation ($z\sim 2$) when galactic winds are at their most important. As our understanding of cosmic ray physics improves it may be fruitful to extend our model to include cosmic rays following existing analytic models for (single-phase) galactic winds driven by cosmic rays \citep[\eg][]{Mao:2018}, with cosmic ray modifications to the cloud-wind interaction informed by simulations \citep[\eg][]{Wiener:2019,Bruggen:2020}.

\emph{Circumgalactic medium impedance/interactions}: Galactic winds do not expand into vacuum, they interact with the surrounding CGM. The interaction of a multiphase wind with a multiphase CGM is a crucial part of how galactic winds regulate the growth of galaxies, and is an active topic of study \citep[\eg][]{Hafen:2020,Stern:2021}. The hot phase of the wind will sweep up and shock the volume filling phase of the CGM, which can heat and eject CGM material as well as lead to the formation of cold shells \citep[\eg][]{Fielding:2017a,Lochhaas:2018}. The expansion of the wind blown bubble will eventually stall when the wind ram pressure no longer exceeds the ambient CGM pressure. Our model provides a reliable estimate for the free-expanding portion of the wind which extends from the galaxy out to the reverse-shock \citep[][]{Weaver:1977}. At the end of the free-expanding portion of the wind the cold phase of the galactic winds will interact with the shocked (hot wind and preexisting CGM) material. This interaction will lead to a deceleration of the clouds that can either be destructive or constructive depending on the cloud properties, much like the acceleration process. Recent semi-analytic studies have investigated how the cloud-CGM interaction can or cannot help explain observations of the cold phase in the context of cosmological inflows \citep{Afruni:2019} and galactic winds \citep{Afruni:2021}. These models, which include only gravity and drag forces on the cloud, provide valuable insights and would be straightforward to extend to include the additional processes of shredding and growth that we have focused on in this work. 

\emph{Hot phase turbulence}: We have assumed throughout that the hot phase is laminar and moves purely in the radial direction. In real galactic winds there can be significant turbulent velocities in the volume-filling phase \citep{Schneider:2020}. This turbulence can change the balance in the competition between shredding and cooling in the turbulent radiative mixing layers that regulate the cloud-wind interaction \citep{Gronke:2021}. This \textit{extrinsic} turbulence may impact the cloud evolution if (i) on the scale of the cloud it exceeds the \textit{intrinsic} turbulence driven by the shear flow, or if (ii) it causes the bulk shear flow to change directions on a timescale shorter than the time it takes the cloud and its wake to equilibriate. In a future work we plan to include the effect of this extrinsic/volume filling phase turbulence by allowing some fraction of the injected energy to be in the form of turbulent kinetic energy which will be advected and decay.

\emph{Cloud coagulation and creation}: In a realistic turbulent galactic wind cold clouds may coagulate and be formed in situ. These processes change the number of clouds at a given radius in a manner not captured by our model, and may be needed to accurately model the cloud mass distribution. Furthermore, a channel for in situ formation via multiphase condensation would enable our model to be extended to the problem of multiphase accretion and galactic precipitation, which is thought to be relevant in halos ranging from dwarfs to clusters \citep[\eg][]{Voit:2017}.

\emph{Time dependence}: Underlying all of our framework is the assumption of steady-state. In many cases this is likely a good assumption, however galactic winds are also known to exhibit a bursty behavior, which may be poorly approximated by our model. In particular, if the wind launching properties vary on timescales longer than $\sim \text{few kpc} / v \approx$ few Myr the assumption of steady state is justified. Furthermore, in cases with large $\etaMcold$ the steady-state solutions fail, which may indicate that a these conditions lead to an inherently time-dependent solution. We plan to address these questions in a future work by using our framework as the basis for a subgrid model in galaxy simulations. 

\emph{Density contrast dependence:} The choices we have made for the cloud mass growth and loss model are by no means the only possibilities that could be broadly consistent with the current simulation results. In particular, the dependence of the mass transfer parameterization on density contrast $\chi$ is worth considering since it has the least stringent constraints from simulations. For example, one could adopt a more elongated cloud, with $\Acool \propto \chi \rcl^2$, or a $\chi$-dependent $\vturb$, such as $\fturb = \chi^{-1/2}$. These choices will change the normalization of the mass transfer rates $\Mdo$ and the critical cloud mass above which $\Mdotcl > 0$, but will not qualitatively change any of the findings. 

\emph{Mach dependence}: One of the most important simplifications to the cloud-wind interaction we made, which will be refined in the future, is the lack of a Mach dependence. The clouds in our multiphase wind model often have highly supersonic relative velocities ($\vrel / \cshot$). Numerical simulations of such high Mach number cloud-wind interactions are challenging, so it remains unclear how the mass, momentum, and energy transfer depends on the Mach number. There is, however, an indication that the characteristic cloud destruction time scale should be lengthened by a factor of $\sqrt{1+\Mach}$ \citep{Scannapieco:2015}.

\emph{Distributed cloud sweep-up in the subsonic region}: The simple volumetric injection of hot material at small radii in our model drives the wind which immediately runs into a thin shell of cold clouds outside of the injection region. This abrupt introduction of the cold clouds into the hot wind neglects the fact that many (if not all) of the cold clouds are initially co-spatial with the supernova remnants that are powering the wind. A more physical approach may be to introduce the clouds distributed uniformly within the same (subsonic) region that the hot material is added. 

\emph{Unexplored dependencies}: In addition to the long list of missing ingredients and model extensions there are also parameter dependencies that we have yet to fully explore. In particular, the structure of multiphase galactic winds has interesting albeit less significant dependence on $\vc$, $Z$, and $\rstar$, that we have not presented for the sake of brevity. We encourage interested readers to explore these dependencies using our code that is available here: \url{https://github.com/dfielding14/MultiphaseGalacticWind/}.


\section{Summary} \label{sec:summary}

This work presents a new analytic model that describes the steady state structure of a spherically symmetric multiphase galactic wind that is comprised of cold clouds and a volume-filling hot component. The novel aspect of our model, which is also the fundamental driver of much of the interesting behavior of the winds, is our treatment of the exchange of mass, momentum, and energy between the hot and cold phase. This exchange is regulated in the turbulent radiative mixing layers that develop between the phases. The transfer rates are set by the competition between turbulence, which shreds the clouds, and cooling, which grows the clouds, with the crucial parameter being the ratio of the mixing time to the cooling time $\xi = \tmix / \tcool$. When $\xi>1$ the cloud is in the growth regime, and when $\xi<1$ the cloud is in the shredding regime. Simulations of turbulent radiative mixing layers and cloud-wind interactions have, recently, illuminated the nature of these competing processes and enabled us to write down intuitive analytic approximations that are ideally suited for use in a galactic wind model. By including this essential and inherently small scale phenomena into a global context we have been able to explore not only the impact of the hot wind on the cold clouds, but the previously under-appreciated back reaction of the cold clouds on the hot wind itself. A schematic summary of our model is shown in \autoref{fig:Schematic}.

In \autoref{sec:volumefilling} we lay the foundation for our model by exploring the general behavior of a hot wind under the influence of arbitrary sources/sinks of mass, momentum, and energy. Limiting cases reproduce well studied cases of galactic wind solutions, driven by centrally injected mass and energy with or without gravity and cooling (see \autoref{sec:review_limit_cases}). We then refine this generic hot wind radial structure formulation to account for the specific case in which material can be either gained from, or lost to, another phase that is moving with a velocity relative to the hot phase (see \autoref{sec:source_terms_relative_velocity}). In addition to the change in mass, momentum, and energy from this transfer, our formulation also accounts for additional forces, such as drag/ram pressure, and additional energy source terms, such as radiative cooling (see \autoref{eq:split_source}). This leads us to the fundamental equations that govern the radial evolution of the hot phase of a galactic wind (see \autoref{eq:velocity_gradient_split_source} to \autoref{eq:entropy_gradient_split_source}). Although these equations for the hot phase velocity $v$, density $\rho$, pressure $P$, and entropy $K$, have up to ten separate terms, each term has a well-defined physical interpretation, which we use to understand what is driving the properties of the wind (see \autoref{fig:Gradients}).

The source terms for the hot phase come from the interaction of the hot phase with the cold clouds that are embedded within it. In \autoref{sec:cloudevo}, we present a framework for modeling the cloud-wind interaction that, although simple, encapsulates the essential behaviors. Specifically, our framework describes the mass transfer, accelerations, and energetics of this interaction. We present a physically motivated parameterization of the mass evolution of an embedded cold cloud that is inspired by and consistent with extensive recent work on the topic (see \autoref{sec:mass_transfer_model} and \autoref{fig:fiducial_cloud_wind}). In developing this parameterization we stress where our specific choices are underconstrained by simulations or oversimplified and can be refined in future work.

Finally, we present the full model for the coevolution of the cold clouds and the hot wind in \autoref{sec:coevolution}. This model allows us to self-consistently describe the radial structure of the hot wind and the properties of the cold clouds for a given set of wind launching parameters. The primary parameters that determine the wind properties are the SFR, hot phase mass loading $\etaM$, hot phase energy loading $\etaE$, cold phase mass loading $\etaMcold$, and the initial mass of cold clouds $\Mcl$.\footnote{Additional parameters include the size of the star forming region $\rstar$, hot phase metallicity, and cold phase metallicity, we, however, keep these fixed at 300 pc, 2 $Z_\odot$, and $Z_\odot$, respectively.} We demonstrate how the presence of cold clouds can dramatically impact the structure of the hot wind and how the clouds themselves evolve as they flow out into the surrounding medium. Some of the notable features of the multiphase wind solutions are as follows.
\begin{enumerate}[i.]
  \setlength{\parskip}{0pt}
  \setlength{\itemsep}{0pt plus 1pt}
   \item We find that the multiphase model presented here naturally leads to galactic winds in which the energy from supernovae is contained within a low-density, high-specific-energy fluid, while cold dense clouds are accelerated and entrained by the turbulent radiative mixing layers that form at their boundaries. This allows the winds to be effective agents of feedback in the circumgalactic medium, providing a mechanism by which star formation in the galactic disk can slow cooling and accretion as part of a global {\it preventive} self-regulation mechanism. In addition, the winds can transport cold clouds to large radii, in agreement with observations\footnote{We note that the amount of mass in the hot phase is relatively small, meaning that it is very challenging for cosmological galaxy-scale simulations (almost all of which adopt fixed mass resolution) to resolve. Such simulations tend to mix the two phases, resulting in a wind which cools much more rapidly than the multiphase structure envisioned here, requiring the winds to be essentially momentum-driven, and therefore highly mass loaded. We demonstrate in the current paper that is not required and that, in real galactic systems, the high-specific energy hot wind does not (necessarily) lose significant amounts of energy so that low-mass loaded winds can effectively transmit energy into the circumgalactic medium.}.
  \item The bi-directional interaction between the hot and cold phases is of critical importance in the model. The acceleration of the cold clouds, which is almost always dominated by the accretion of cooled wind material onto the clouds during the turbulent radiative mixing layer entrainment process (see \autoref{fig:Gradients}), comes at the expense of the hot phase velocity, which can be reduced by a factor $>3$ relative to an otherwise identical wind without the cold clouds. Furthermore, the fact that cold clouds are predominantly accelerated by the exchange of mass with the hot phase explains the close correspondence between the degree of metal mixing and the clouds velocity (see \autoref{fig:Case_I_low_eta} and \autoref{fig:Case_I_high_eta}).
  \item The thermalization of the relative kinetic energy as cloud material is mixed into the hot wind leads to a significant heating of the wind. When cooling is unimportant in the hot phase, which happens when the hot phase mass loading is low ($\etaM \sim 0.1$), this heating increases the hot wind entropy, and results in a flatter temperature profile (see \autoref{fig:Case_I_low_eta} and \autoref{fig:Gradients}). Flatter temperature profiles have been found in high resolution galactic wind simulations \citep{Fielding:2017b, Schneider:2020}, as well as in observations \citep{Lopez:2020}.
  \item In galactic winds with a low hot phase density, which occurs when either $\etaM$ and/or SFR is small, cold clouds tend to lose mass (see \autoref{fig:Case_I_low_eta}). This is true even with the most massive cold clouds $\Mcl = 10^6 \ \Msun$, which start out in the growing regime ($\xi > 1$) but quickly transition to being shredded ($\xi < 1$). This is because the cooling rate in the turbulent radiative mixing layers drops precipitously as the pressure drops. As the clouds are shredded the mass flux in the cold phase drops while the hot phase becomes increasingly mass loaded. In cases in which there is a large initial cold mass loading factor and the clouds are rapidly shredded ($\xi \lesssim 0.1$) the rapid addition of material into the hot phase can cause the wind to \emph{fail} by driving the Mach number down below unity. Winds with small $\etaM$ and/or low SFRs are able to sustain higher total mass loading ($\etaMtot = \etaM + \etaMcold$), but are likely to become increasingly dominated by the hot phase at larger distances (see \autoref{fig:Contours}, \autoref{fig:Contours_SFR}, and \autoref{fig:eta_M_edot_SFR}).
  \item On the other hand, in galactic winds with a higher hot phase density, which occurs when either $\etaM$ and/or the SFR is large, cold clouds tend to gain mass because of the short cooling times in their turbulent radiative mixing layers (see \autoref{fig:Case_I_high_eta}). As a result of cloud growth, the mass flux in the hot phase drops and the density profile steepens. In these rapid cloud-growth systems the hot phase energy flux is dramatically reduced. In some cases the hot wind loses all of its thermal energy and cools down to the clouds temperature ($10^4$ K) through the combination of cooling in the mixing layer and in the volume filling phase. These winds have in effect been \emph{poisoned} by sweeping up too many cold clouds. This processes limits the maximum cold phase mass flux, and as a result winds with large $\etaM$ and/or high SFRs are not able to sustain as high of a total mass loading factor ($\etaMtot = \etaM + \etaMcold$), but are likely to remain multiphase (or even predominantly cold) to large distances (see \autoref{fig:Contours}, \autoref{fig:Contours_SFR}, and \autoref{fig:eta_M_edot_SFR}).
\end{enumerate}
\vspace{-0.5\topsep}

The model we have presented provides a flexible, intuitive, and self-consistent framework to understand the nature of galactic winds. Our analytic approach is an ideal complement to simulations because it reproduces many of the key processes in both small and large scale simulations while being easily extensible and incurring essentially no computational cost, making it well-suited for comparison to diverse observations. This first incarnation of the model sacrifices some physical realism for the sake of simplicity, but there is a clear path forward to include extensions. These extensions could include improved parameterizations, the inclusion of additional processes, a full accounting for the phase structure necessary for detailed observational comparisons, and the model's implementation as a cosmological simulation subgrid model. We suspect that these extensions will only strengthen the key finding of this work that accounting for the full galactic wind phase structure is crucial for understanding the flow of mass, momentum, and energy into and out of galaxies which regulates their formation over cosmic time.

\acknowledgements 
We have derived pleasure and profit from discussing the ideas in this paper with many colleagues. In particular we would like to thank Matthew Smith, Eve Ostriker, Alberto Bolatto, Eliot Quataert, Chang-Goo Kim, John Forbes, Matthew Abruzzo, Anthony Chow, Zirui Chen, Evan Schneider, Max Gronke, Peng Oh, and the SMAUG team for many useful discussions and suggestions. We thank Lucy Reading-Ikkanda for help making \autoref{fig:Schematic}. DBF is supported by the Simons Foundation through the Flatiron Institute. GLB acknowledges financial support from the Simons Foundation, NSF (grant OAC-1835509), and NASA (grant NNX15AB20G). We thank KITP for hosting the Fundamentals of Gaseous Halos workshop (supported by the NSF under Grant No. NSF PHY-1748958) during which some of the key ideas of this work were formulated.

\emph {Code availability:} A basic implementation of the code used to solve for the steady state structure of a multiphase wind that was used to create all of the figures in this paper is available here: \url{https://github.com/dfielding14/MultiphaseGalacticWind/}.

\emph {Software:} \texttt{matplotlib} \citep{matplotlib}, \texttt{scipy} \citep{scipy}, \texttt{numpy} \citep{numpy}, \texttt{ipython} \citep{ipython}, \texttt{CMasher} \citep{CMasher}.

\bibliography{references}

\end{document}